\documentclass[aps,
prd,
onecolumn,
amsmath,amssymb,
showkeys,
nofootinbib,
floatfix,
superscriptaddress,
]{revtex4-2}

\usepackage{hyperref}
\hypersetup{
    colorlinks=true,
    linkcolor=red,
    filecolor=magenta,      
    urlcolor=blue,
    citecolor = teal,
}

\usepackage{graphicx}
\usepackage[dvipsnames]{xcolor}
\usepackage[normalem]{ulem}

\newcommand{\replaceA}[2]{{{\color{BrickRed}{#1}}{\color{NavyBlue}{\ifmmode\text{\sout{\ensuremath{#2}}}\else\sout{#2}\fi}}}}

\newcommand{\replaceV}[2]{{{\color{RedViolet}{#1}}{\color{BlueViolet}{\ifmmode\text{\sout{\ensuremath{#2}}}\else\sout{#2}\fi}}}}

\newcommand{\replaceJ}[2]{{{\color{RedOrange}{#1}}{\color{OliveGreen}{\ifmmode\text{\sout{\ensuremath{#2}}}\else\sout{#2}\fi}}}}

\newcommand{\changeq}[3]{{\color{red} \ifmmode\text{\sout{\ensuremath{#1}}}\else\sout{#1}\fi}{\color[rgb]{0.56,0.0,1.0} #2}{\color{blue}[#3]}}
\newcommand{\del}[1]{ {\color{violet}{\ifmmode\text{\sout{\ensuremath{#1}}}\else\sout{#1}\fi}} }

\usepackage{dcolumn}		

\usepackage{MnSymbol}

\newcommand{\bs}{\boldsymbol}
\DeclareMathOperator{\Tr}{Tr}

\newcommand{\beq}{\begin{eqnarray}}
\newcommand{\eeq}{\end{eqnarray}}

\allowdisplaybreaks

\begin{document}

\title{%
Spatial confinement-deconfinement transition in accelerated gluodynamics within lattice simulation
}

\author{Victor~V.~Braguta} 
\email{vvbraguta@theor.jinr.ru}
\affiliation{Bogoliubov Laboratory of Theoretical Physics, Joint Institute for Nuclear Research,  Dubna 141980, Russia}
\author{Vladimir~A.~Goy} 
\email{vovagoy@gmail.com}
\affiliation{Pacific Quantum Center, Far Eastern Federal University, Vladivostok 690922, Russia}
\affiliation{Institute of Automation and Control Processes, Far Eastern Branch, Russian Academy of Science, 5 Radio Str., Vladivostok 690041, Russia}
\author{Jayanta~Dey}
\email{jayanta@theor.jinr.ru}
\affiliation{Bogoliubov Laboratory of Theoretical Physics, Joint Institute for Nuclear Research,  Dubna 141980, Russia}
\author{Artem~A.~Roenko}
\email{roenko@theor.jinr.ru}
\affiliation{Bogoliubov Laboratory of Theoretical Physics, Joint Institute for Nuclear Research,  Dubna 141980, Russia}

\date{\today}

\begin{abstract}
In this work we investigate the influence of weak acceleration on the confinement-deconfinement phase transition in gluodynamics. Our study is carried out within lattice simulation in the  comoving reference frame of  accelerated observer which is parameterized by the Rindler coordinates.
We find that finite temperature confinement-deconfinement phase transition turns into spatial crossover in the Rindler spacetime. In other words, spatially separated confinement and deconfinement phases can coexist in the Rindler spacetime within certain intervals of temperature and acceleration.
We determine the position of the boundary between the phases as a function of temperature for several accelerations and find that it can be described by the Tolman-Ehrenfest law with rather good accuracy although a minor deviation takes place. 
Moreover, the critical temperature of the system in the weak acceleration regime is found to remain unchanged as that of the standard homogeneous gluodynamics.
Our results imply that the spatial confinement-deconfinement transition might take place  in the vicinity of the Schwarzschild black hole horizon. 
\end{abstract}

\maketitle

\section{Introduction}\label{sec:Intro}

\begin{figure*}[t]
    \centering
    \includegraphics[width=1.0\linewidth]{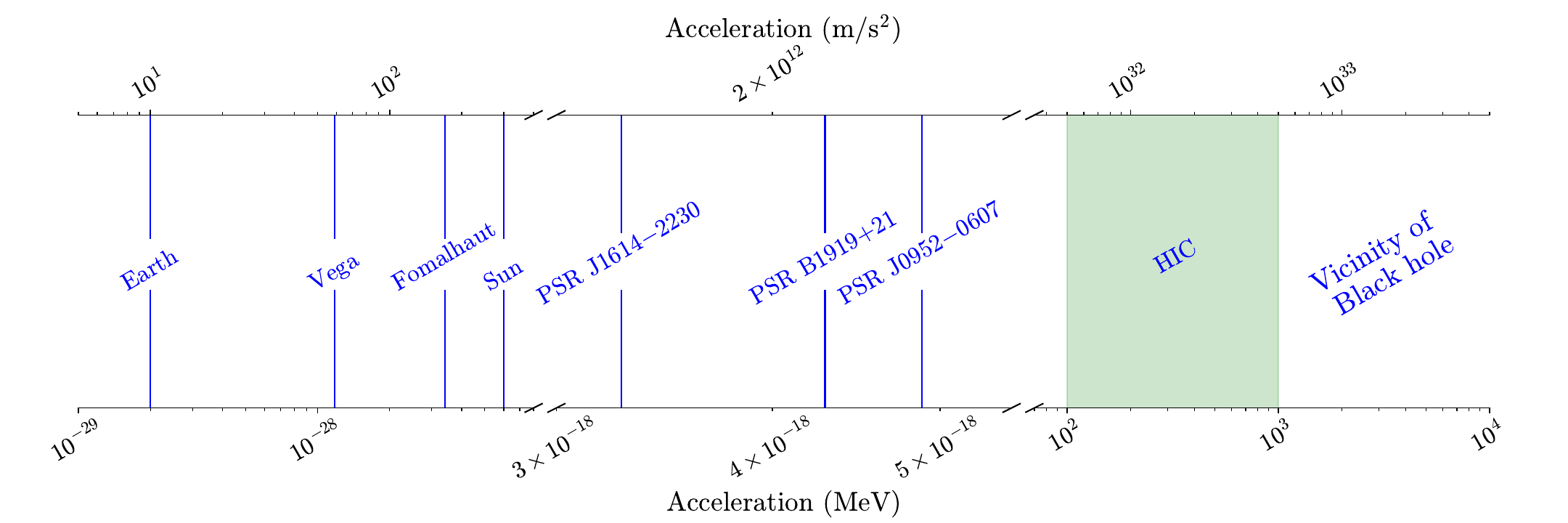}
    \caption{%
    Comparison of characteristic values of gravitational acceleration near various astrophysical objects and in heavy-ion collision experiments~(HIC). In particular, as an example of such objects we chose the stars: Vega, Fomalhaut, Sun and the neutron stars: PSR J1614-2230,  PSR B1919+21, PSR J0952-0607. For the accelerations which can be reached in HIC experiments we shade the region from 100~MeV to 1~GeV~\cite{Prokhorov:2025vak, Kharzeev:2005iz}. Classically arbitrary large acceleration can be achieved in the vicinity of black hole horizon. 
    }
    \label{fig:acceleration}
\end{figure*}

Modern heavy-ion collision~(HIC) experiments offer a unique opportunity to study the behavior of quark-gluon matter under extreme conditions, including high temperatures, intense magnetic fields, relativistic rotation, and enormous accelerations. This paper investigates the effects of acceleration. The characteristic accelerations that might be achieved in these collisions can reach values as high as $\sim 0.1-1$~GeV according to the estimations in Refs.~\cite{Prokhorov:2025vak, Kharzeev:2005iz}.
Such a large acceleration is  expected to profoundly influence Quantum Chromodynamics (QCD) properties. In addition, comparing  these values with  accelerations of free fall near different astrophysical objects (see Fig.~\ref{fig:acceleration}), one can draw a conclusion that the accelerations achieved in HIC can only be  met in the vicinity of a black hole horizon. So, HIC experiments might be used to study quantum effects in gravity or even quantum gravity. 

A collision of heavy nuclei followed by rapid expansion of the created plasma fireball is a very complex non-equilibrium process. If we consider magnetic field created in such a collision, it is neither homogeneous in space 
nor constant in time. However, it has become commonplace to investigate QCD properties within a simplified model: QCD in thermal equilibrium affected by a homogeneous and time independent magnetic field. Similarly,  acceleration in HIC depends on time and spatial coordinates. However, as a study of a simplified model, one can explore QCD properties in the uniformly accelerated reference frame, which is parameterized by the Rindler coordinates~\cite{Crispino:2007eb, Mukhanov:2007zz}. 

A defining feature of a uniformly accelerated reference frame is the emergence of a Rindler event horizon~\cite{Rindler:1966zz}. Since particles beyond this horizon cannot reach the accelerated observer, the observer perceives the Minkowski vacuum as a thermal bath of particles at the Unruh temperature~\cite{Unruh:1976db}
\begin{equation}
T_U = \hbar \frac { \alpha} {2\pi k c}\,,
\label{eq:unruh_temperature}
\end{equation}
where $\alpha$ is the acceleration\footnote{To avoid confusion with lattice spacing $a$, in our paper acceleration is designated as $\alpha$, which differs from the commonplace designation for the thermal acceleration $\alpha$ (see, for instance, Ref.~\cite{Becattini:2015nva}).},
$\hbar$ is the Planck constant, $k$ is the Boltzmann constant, and $c$ is the speed of light.\footnote{In what follows it is assumed that $\hbar=c=k=1$.}
The formula for the Unruh temperature represents an intriguing interplay between gravity, quantum field theory, and statistical physics, which continues to attract the interest of various theoretical groups (see, for instance,~\cite{Prokhorov:2019sss, Prokhorov:2023dfg, Khakimov:2023emy, Ambrus:2023smm, Chernodub:2024wis, Prokhorov:2025vak, Akhmedov:2021agm, Diakonov:2023hzg, Akhmedov:2023qmg, Selch:2023pap, Tsegelnik:2026ulr, Bordag:2025zbt, Becattini:2015nva, Becattini:2017ljh, Becattini:2020qol, Palermo:2021hlf}).

A question thoroughly discussed in the literature is the restoration of broken symmetries under sufficiently large acceleration~\cite{Ohsaku:2004rv, Kharzeev:2005iz, Ebert:2006bh, Castorina:2012yg, Takeuchi:2015nga, Benic:2015qha, Dobado:2017xxb, Casado-Turrion:2019gbg, Kou:2024dml, Chernodub:2025ovo, Zhu:2025pxh}.
This is particularly interesting in QCD, where the chiral symmetry is broken in the low-temperature phase, resulting in a nonzero chiral condensate. Thermal fluctuations diminish the chiral condensate and lead to the restoration of chiral symmetry at a sufficiently high temperature. A similar phenomenon is hypothesized for a uniformly accelerated observer, who could, in principle, use the chiral condensate as a thermometer to measure the Unruh temperature of the vacuum. Based on this idea it is reasonable to assume that there exists critical acceleration for which the chiral symmetry is restored. This pattern of the chiral symmetry breaking-restoration transition is supported by a number of studies~\cite{Ohsaku:2004rv, Kharzeev:2005iz, Ebert:2006bh, Castorina:2012yg, Casado-Turrion:2019gbg, Kou:2024dml}.

In contrast, the authors of Refs.~\cite{Unruh:1983ac, Salluce:2024jlj} argue that an accelerated observer cannot detect such a phase transition in the Minkowski vacuum. The disagreement between these approaches is beyond the scope of our present work, as we are going to focus specifically on phase transitions within thermodynamically equilibrated, accelerated quark-gluon matter. 
In connection to this issue it is interesting to mention the results of Ref.~\cite{Chernodub:2025ovo}. As was shown in this paper, uniform acceleration of matter produces an effect similar to cooling and  enhances the effect of spontaneous symmetry breaking. 

This paper addresses the confinement-deconfinement phase transition in weakly accelerated ($\alpha \ll \Lambda$) gluodynamics.\footnote{In our paper the characteristic energy scale of gluodynamics  designated as $\Lambda$, which can be estimated as $\Lambda \sim 200$~MeV~\cite{Luscher:2010iy}.}
While the influence of weak acceleration on this transition was first studied in Ref.~\cite{Chernodub:2024wis}, their approach utilized the Tolman-Ehrenfest (TE) law~\cite{Tolman:1930ona} to reduce the effect of acceleration to a properly adjusted temperature gradient. In contrast,  our work employs an alternative and straightforward approach by studying gluodynamics directly within Rindler coordinates~\cite{Crispino:2007eb, Mukhanov:2007zz}, which parametrizes a uniformly accelerated reference frame. To account for the strong interactions in gluodynamics, we will apply lattice simulation.

It is worth noting that lattice simulations have recently been applied to study the confinement-deconfinement phase transition in  rotating reference frame~\cite{Yamamoto:2013zwa, Braguta:2020biu, Braguta:2021jgn, Braguta:2022str, Yang:2023vsw, Braguta:2023kwl, Braguta:2023iyx, Braguta:2024zpi}. A remarkable result, discovered in papers~\cite{Braguta:2023iyx, Braguta:2024zpi}, is the existence of a mixed, inhomogeneous phase where confinement and deconfinement coexist while spatially separated, within certain intervals of temperature and angular velocity. 
In Ref.~\cite{Braguta:2024zpi}, it was shown that this phenomenon can be explained by a modification of the gluon action due to the effective gravitational field induced by rotation. By analogy, it is reasonable to assume that the gravitational field in Rindler spacetime might similarly modify the gluon action, potentially giving rise to an inhomogeneous phase in the accelerated case as well. The present work will investigate this possibility.

This paper is organized as follows.
In Section~\ref{sec:Acceleration} we review the properties of the accelerated motion and the Rindler spacetime which parameterize the comoving reference frame of uniformly accelerated observer.
In Section~\ref{sec:Theory} we consider the properties of gluodynamics in the Rindler spacetime and build its lattice version. In addition, in this section lattice set up and measured observables are described. Section~\ref{sec:Results} is devoted to the results of this work. In particular, we study the properties of spatial confinement-deconfinement phase transition and  the position of the boundary between confinement and deconfinement phases. Finally in Section~\ref{sec:Conclusions} we summarize our results and draw a conclusion. Relevant theoretical and technical details are provided in the Appendix.

\section{Comoving frame of accelerated observer} \label{sec:Acceleration}

\subsection{Accelerated motion}
\label{sec:Acceleration_Observer}

In this section, we are going to review construction of the comoving frame for an accelerated observer~--  the Rindler spacetime.\footnote{Here we follow Ref.~\cite{Mukhanov:2007zz}}
To begin, let us consider the trajectory of an accelerated observer in 1+1 dimensional Minkowski space with the temporal coordinate $\xi^0$ and the spatial coordinate $\xi^1$. To simplify the formulas we are going to use the lightcone coordinates 
\begin{equation} 
x^{0} = u = \xi^0-\xi^1,\qquad
x^{1} = v = \xi^0+\xi^1 \,,
\end{equation}
for which the interval can be written as $ds^2 =  (d\xi^0)^2 - (d\xi^1)^2 = dx^{\mu} dx_{\mu} =du\,dv$. Notice that $x_0=1/2 v$, $x_1=1/2 u$ and the Lorentz transformation that preserves the interval is $u \to \gamma \cdot u$, $v \to 1/\gamma \cdot v$.

To parameterize the trajectory one can use the proper time $\tau$. With this parameterization the 2-velocity of the observer has the form 
\begin{equation}
  u^{\mu} = \frac {d x^{\mu}} {d \tau} = \dot x^{\mu}\,,
\end{equation}
and obeys the following normalization condition:
\begin{equation}
  u^{\mu} u_{\mu} = \dot u(\tau)  \dot v(\tau) = 1\,.
  \label{eq:normalization}
\end{equation}
The meaning of constant acceleration can be clarified if one passes to the instantly comoving inertial frame. In this reference frame, the observer is at rest and experiences constant acceleration $d^2 \xi^1 / d \tau^2 = \alpha =const$. The last relation can be written in the covariant form valid for arbitrary reference frame:
\begin{equation}
a^{\mu} a_{\mu} = {\ddot u(\tau)}  \ddot v(\tau) = -\alpha^2 \,.
\label{eq:normalization_acceleration}
\end{equation}
By extracting $\dot u(\tau)$ from Eq.~\eqref{eq:normalization}, $\dot u = 1 / \dot v$, and substituting it into Eq.~\eqref{eq:normalization_acceleration}, one gets the differential equation for the coordinate $v(\tau)$,
\begin{equation}
\left(  \frac {\ddot v} {\dot v}  \right)^2= \alpha^2\,.
\end{equation}
The solution of this equation gives the coordinate $v(\tau)$ as a function of time, while the second coordinate, $u(\tau)$, can be expressed using the normalization condition~\eqref{eq:normalization_acceleration},
\begin{equation}
v(\tau) = \frac A {\alpha} e^{\alpha \tau} + B\,,\qquad
u(\tau) = - \frac 1 {A\alpha} e^{-\alpha \tau} + C\,, 
\end{equation}
where $A,\, B,\, C$ are integration constants. Applying Lorenz transformation and shifting the origin, the trajectory of the accelerated observer can be written as 
\begin{equation}
u(\tau) = - \frac 1 {\alpha} e^{- \alpha \tau}\,, \qquad
v(\tau) = \frac 1 {\alpha} e^{ \alpha \tau}\,.
\label{eq:trajectory}
\end{equation}
In the Minkowski coordinates the trajectory is  
\begin{equation}
\xi^0(\tau) = \frac {v+u}2  = \frac 1 {\alpha} \sinh \alpha \tau\,, \qquad
\xi^1(\tau) = \frac {v-u} 2  =\frac 1 {\alpha} \cosh \alpha \tau\,.
\end{equation}
These formulas show that the accelerated observer moves along a branch of hyperbola $(\xi^1)^2 - (\xi^0)^2 = 1/\alpha^2$. In the chosen coordinate system the observer decelerates in the time region $\xi^0 \in (-\infty, 0)$ reaching the coordinate $\xi^1=1/\alpha$ at the moment $\xi^0=0$ and then accelerates in the region $\xi^0 \in (0,\infty)$. 

\subsection{The Rindler spacetime}
\label{sec:Acceleration_Rindler}

Further, let us proceed to the construction of the coordinates $(\eta^0,\eta^1)$ which parameterize  the comoving frame of uniformly accelerated observer.  In order to do this we impose the following conditions:
\begin{enumerate}
\item The observer in this reference frame is at rest in the spatial coordinate origin $\eta^1=0$.
\item The temporal coordinate for the observer coincides with the proper time $\eta^0 = \tau$.
\item The coordinates $(\eta^0,\eta^1)$ determine the conformally flat metric 
\begin{equation}
ds^2 = \Omega^2(\eta^0, \eta^1) \left( (d \eta^0)^2 - (d \eta^1)^2 \right)\,, 
\end{equation}
where $\Omega(\eta^0, \eta^1)$ is a function of the coordinates, which will be specified below. 
\end{enumerate}
Similarly to the previous section, instead of the $(\eta^0,\eta^1)$ coordinates it is convenient to use lightcone coordinates
\begin{equation}
\tilde u = \eta^0 - \eta^1\,, \qquad \tilde v = \eta^0 + \eta^1\,.
\end{equation}
Evidently, at the observer's  worldline the lightcone coordinates are $\tilde u(\tau) = \tilde v(\tau) = \tau$ and $\Omega(\tilde u=\tau, \tilde v=\tau)=1$.  

The interval can be expressed either through the lightcone coordinates in Minkowski spacetime, $(u,v)$, or through the $(\tilde u, \tilde v)$ coordinates:
\begin{equation}
ds^2 = du\, dv = \Omega^2(\tilde u, \tilde v ) d \tilde u\, d \tilde v\,,
\label{eq:interval}
\end{equation}
which are related by some coordinate transformation, $u=u(\tilde u, \tilde v)$ and $v=v(\tilde u, \tilde v)$.
Notice, however, that an arbitrary transformation violates the conformal property of the metric~\eqref{eq:interval} due to the appearance of the $d \tilde u^2$, $ d \tilde v^2$ terms in the expression for the interval. The only possibility to respect this property is that $u=u(\tilde u)$,\,  $v=v(\tilde v)$ or $u=u(\tilde v)$,\,  $v=v(\tilde u)$. 
In the following, we are going to use the former option, without loss of generality. 

The explicit form of the coordinate transformation can be determined on the trajectory~\eqref{eq:trajectory}. In order to do this, let us consider the equality 
\begin{equation}
\frac {d u (\tau)} {d \tau} = \frac {d u (\tilde u)} {d \tilde u}
\frac {d \tilde u (\tau)} {d \tau}\,.
\label{eq:derivative}
\end{equation}
Due to the second imposed condition, we have ${d \tilde u (\tau)}/{d \tau}=1$, whereas the left-hand side of this equation, ${d u (\tau)} / {d \tau}$, can be found from the observer's trajectory~\eqref{eq:trajectory}, 
\begin{equation}
\frac {d u (\tau)} {d \tau} = e^{-\alpha \tau} = -\alpha u(\tilde u)\,.
\end{equation}
Substituting all ingredients to the equation~\eqref{eq:derivative}, we get the differential equation 
\begin{equation}
\frac {d u(\tilde u)} { d \tilde u} = - \alpha u(\tilde u)\,,
\end{equation}
which can be readily solved  
\begin{equation}
u=C_1 e^{-\alpha \tilde u}\,,
\end{equation}
where $C_1$ is a constant. Similarly, 
\begin{equation}
v=C_2 e^{\alpha \tilde v}\,.
\end{equation}
The condition $\Omega(\tilde u=\tau, \tilde v=\tau)=1$ imposes the restriction on the integration constants: $\alpha^2 C_1 C_2 = -1$. If one chooses $C_1=-C_2$, the coordinate transformation becomes 
\begin{equation}
 u=-\frac 1 {\alpha} e^{-\alpha \tilde u},\qquad v=\frac 1 {\alpha} e^{\alpha \tilde v}, 
\end{equation}
leading to the following interval in the lightcone coordinates:
\begin{equation}
ds^2= e^{\alpha (\tilde v - \tilde u)} d \tilde u\, d \tilde v\,.
\end{equation}
Using formulas derived above, one can write the Minkowski coordinates $(\xi^0,\xi^1)$ in terms of the Rindler coordinates $(\eta^0, \eta^1)$ as
\begin{equation}
\xi^0(\eta^0, \eta^1) = \frac 1 {\alpha} e^{\alpha \eta^1} \sinh \alpha \eta^0,\qquad 
\xi^1(\eta^0, \eta^1) = \frac 1 {\alpha} e^{\alpha \eta^1} \cosh \alpha \eta^0.
\label{eq:coord_transformation}
\end{equation}
Finally, the interval for uniformly accelerated observer is
\begin{equation}
ds^2 = e^{2 \alpha \eta^1} \left (  (d\eta^0)^2 - (d\eta^1)^2 \right )\,,
\label{eq:Rindelr_interval}
\end{equation}
which describes the Rindler spacetime. Notice that the Rindler coordinates are related to the Minkowski ones by the coordinate transformation~\eqref{eq:coord_transformation}, so the Rindler spacetime is equivalent to the Minkowski spacetime and has zero curvature. The line $\xi^0=\xi^1$ corresponds to the horizon for the accelerated observer, since no signal above this line can reach the observer. 

\begin{figure}[t]
    \centering
    \includegraphics[width=0.40\linewidth]{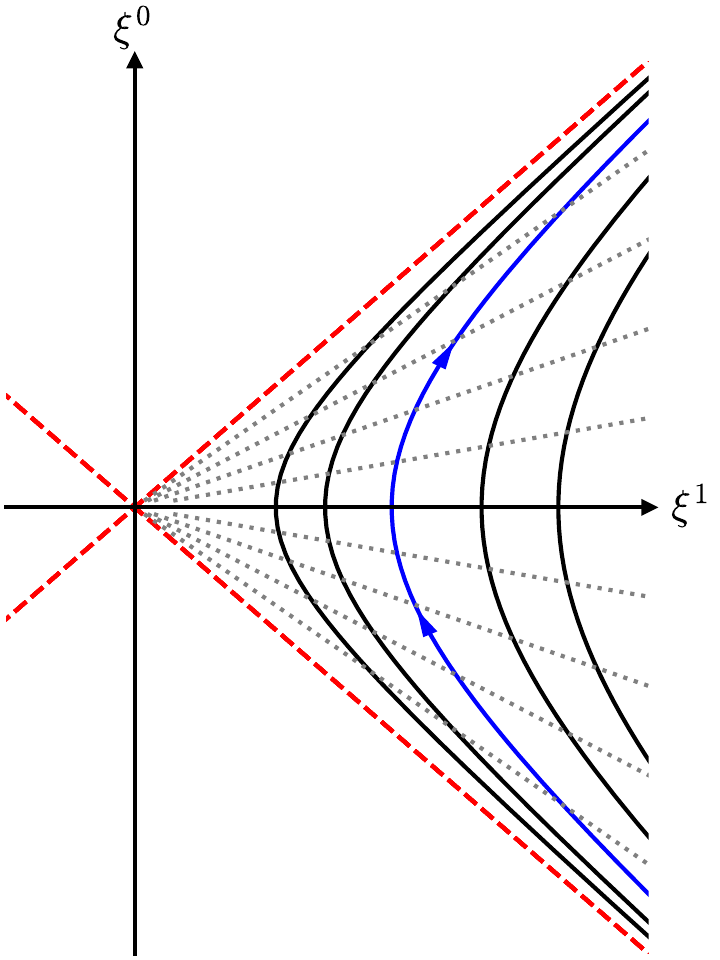}
    \caption{%
    The coordinate system of a uniformly accelerated observer in Minkowski spacetime (see formulas (\ref{eq:coord_transformation})). The hypersurfaces $\eta^1=const$ are depicted by the black solid hyperbolas $(\xi^1)^2-(\xi^0)^2=e^{2\alpha\eta^1}/\alpha^2$.
    The trajectory of the observer is represented by the blue hyperbola $(\xi^1)^2-(\xi^0)^2=1/\alpha^2 $.
    The hypersurfaces $\eta^0=const$ are depicted by black dotted rays $\xi^0/\xi^1 = \tanh \alpha \eta^0$. The red dashed lines show the lightcone and boundary for the space which can be parameterized by the coordinates (\ref{eq:coord_transformation}).
    }
    \label{fig:Minkowski}
\end{figure}

In Fig.~\ref{fig:Minkowski} we show the lines of constant $\eta^0$ and $\eta^1$ in the Minkowski spacetime. From this plot it is seen that although the coordinates $\eta^0, \eta^1$ vary from $-\infty$ to $+\infty$ they cover only a quarter of the Minkowski spacetime, which is called the right Rindler wedge~\cite{Crispino:2007eb}. Notice, however, in our work we consider the thermodynamic equilibrium in a finite volume, and the Rindler coordinates are sufficient for our study. 

The equivalence principle is a fundamental concept in general relativity stating that the effects of gravity are locally indistinguishable from the effects of acceleration. From this perspective the Rindler spacetime is equivalent to the vicinity of the black hole horizon~\cite{Crispino:2007eb}. 
To show this let us consider the Schwarzschild metric in 1+1 dimensions,\footnote{It is assumed that the gravitational constant $G=1$.}
\begin{equation}
ds^2= \left ( 1 -  {2M} / r \right ) dt^2 -  \left ( 1 -  {2M} / r \right )^{-1} dr^2\,.
\label{eq:Schwarzschild}
\end{equation}
By introducing the new coordinate $\rho^2 = 8M(r - 2M)$ near the horizon, $r\simeq 2M$, the Schwarzschild interval~\eqref{eq:Schwarzschild} simplifies to
\begin{equation}
ds^2= \left ( \frac {\rho} {4M} \right )^2 dt^2 -   d\rho^2.
\label{eq:Schwarzschild1}
\end{equation}
One can see that the last expression is equivalent to the Rindler interval~(\ref{eq:Rindelr_interval}) with the associations $t \sim 4Ma \eta^0$ and $\rho \sim  e^{a\eta^1}/a$. For this reason we believe that the results obtained in this paper can be applied to the investigation of gluodynamics/QCD properties near the black hole horizon. Here we would like to mention that the first Monte Carlo study of scalar theory in the background of a classical black hole was carried out in paper~\cite{Benic:2016kdk}.

\section{Lattice formulation of accelerated gluodynamics}\label{sec:Theory}

\subsection{Gluodynamics in the Rindler space}\label{sec:Theory_Rindler}

Let us return to the 3+1 dimensional space. From this point onward, we assume that the acceleration is directed along the $z$-axis and $\alpha>0$.  In this case the Rindler spacetime can be parameterized by the following coordinates: the temporal coordinate $t$ associated with $\eta^0$ of the previous section, two coordinates $x,y$ that are transverse to the acceleration. What concerns the direction of the acceleration,  instead of the $\eta^1$ we are going to use the $z$ coordinate defined as $e^{\alpha \eta^1} = 1 + \alpha z$, for which  the observer is located at $z=0$. Using this setting one can readily write the interval in the 3+1 dimensional Rindler space
\begin{equation}
ds^2 = g_{\mu \nu} dx^{\mu} d x^{\nu} =(1+{\alpha} z)^2 dt^2 - dx^2 - dy^2 -dz^2 \,,
\label{eq:rindler}
\end{equation}
where $dx^{\mu}=(dt, dx, dy, dz)$. 

To build the partition function as a path integral one considers the evolution of the system in the imaginary time $\tau = i t$. The interval~(\ref{eq:rindler}) in this case transforms to
\begin{equation}
ds^2 = (g_E)_{\mu \nu} dx_E^{\mu} dx_E^{\nu}=- (1+{\alpha} z)^2 d\tau^2 - dx^2 - dy^2 -dz^2\,,
\label{eq:Rindler_metric}
\end{equation}
where $dx_E^{\mu}=(dx, dy, dz, d \tau)$. Hereafter, the subscript $E$ stands for Euclidean space.

The partition function for gluodynamics in the Rindler space can be written as the integral over gluon degrees of freedom 
\begin{equation}
\mathcal{Z} = \int \! D A \exp {\left ( -S_E \right )}\,,
\end{equation}
where $S_E$ is the action of gluon fields which has the following form:
\begin{equation}
 S_E = \frac 1 {4 g_{YM}^2} \int \! d^4 x \sqrt g_E g_E^{\mu \nu} g_E^{\alpha \beta} F^a_{\mu \alpha} F^a_{\nu \beta} =  \frac 1 {4 g_{YM}^2} \int d^4 x \left ( \frac 1  {1 +{\alpha} z} {\left({\bs E}^{a} \right)^2} + (1+{\alpha} z) \left({\bs H}^{a} \right)^2
 \right ) \,, 
 \label{eq:SE_continuum}
\end{equation}
where $g_{YM}$ is the strong coupling constant, $F_{\mu \nu}^a$ denotes the gluon field strength tensor, $\bs E^a$ and $\bs H^a$ are chromoelectric and chromomagnetic gluon fields, respectively, and the Rindler metric $(g_E)_{\mu\nu}$ can be extracted from the expression for the interval~\eqref{eq:Rindler_metric}.

Now, a few comments are in order. First, in the path integral formulation of partition function the imaginary time $\tau$ is a compact variable which varies in the region $\tau \in (0, 1/T_0)$. Here, $T_0$ is the temperature at the position of the observer, i.e. at the point with $z=0$.
Next, an important property of the metric~(\ref{eq:Rindler_metric}) is the singularity, $g_{00}(z_h)=0$, which determines the position of the horizon, $z_h=-1/{\alpha}$. Finally, looking to the interval~(\ref{eq:Rindler_metric}) one might assume that the coordinate of the observer $z=0$ stand out from the other coordinates, but this is not the case. One can put the observer to an arbitrary point, say $z_1$. In this case, the following change of variables,  
\begin{align}
{\alpha}_1 & = \frac {\alpha} {1+{\alpha} z_1 }\,, \\ 
d\tau_1 & = (1+{\alpha} z_1) d \tau\,,
\label{eq:TE}
\end{align}
preserves the form of the interval with new parameters: 
\begin{equation}
ds^2 = - (1+{\alpha}_1 (z - z_1) )^2 d\tau_1^2 - dx^2 - dy^2 -dz^2\,.
\label{eq:metric_z1}
\end{equation}
The equation~(\ref{eq:TE}) implies that the variable $\tau_1$ belongs to the interval $\tau_1 \in (0, 1/T_1)$, where  $1/T_1 = (1+{\alpha} z_1)/T_0$, i.e. $T_1 = T_0/(1+{\alpha} z_1)$. Thus we conclude that the observer at the position $z_1$ experiences the acceleration $\alpha(z_1)$ and the temperature $T(z_1)$, which are given by the formulas
\begin{align}
\label{eq:acceleration_z1}
{\alpha}(z_1) & = \frac {\alpha} {1+{\alpha} z_1 }\,, \\
{T}(z_1) & = \frac {T_0} {1+{\alpha} z_1 }\,.
\label{eq:temperature_z1}
\end{align}
In homogeneous system the temperature in thermodynamic equilibrium does not depend on the coordinate. However, for system in an external gravitational field the equilibrium temperature obeys the Tolman-Ehrenfest law~\cite{Tolman:1930ona}, which states that $\sqrt {g_{00}(\bs r)}\cdot T(\bs r) = const$. 
For the Rindler metric~\eqref{eq:rindler}, the $g_{00}$-component is $g_{00} = (1+\alpha z)^2$, and the Tolman-Ehrenfest law takes the form of $T(z)=const/(1+\alpha z)$.
So, the equation (\ref{eq:temperature_z1}), which follows from the transformation to another observer position, expresses the Tolman-Ehrenfest law for the Rindler spacetime. 

An interesting consequence that can be drawn from the Eqs.~\eqref{eq:acceleration_z1} and~\eqref{eq:temperature_z1} is that when approaching to the horizon, the acceleration and the temperature increase. Both parameters are singular at the horizon $z_h=-1/\alpha$. On the contrary, if one moves away from the horizon, the temperature as well as the acceleration drop and at large distances we return to the Minkowski vacuum of gluodynamics. In our paper we are going to focus mostly on the case of weak accelerations $\alpha \zeta \ll 1$,  where $\zeta$ is the correlation length that can be estimated as the inverse characteristic energy scale in gluodynamics $1/\zeta \sim \Lambda$. Thus, in our paper, we consider accelerations $\alpha \ll \Lambda$. 
In addition, it is assumed that the lattice volume under investigation is far from the horizon, i.e. $\alpha(z) \ll \Lambda$ for all lattice coordinates.

The property described by the formulas~(\ref{eq:acceleration_z1}) and~(\ref{eq:temperature_z1}) allows one to anticipate the behavior of the gluodynamics in Rindler spacetime for weak accelerations $\alpha$. To do this, let us consider the gluon field from the point of the observer located at $z=z_1$. Applying the formulas (\ref{eq:acceleration_z1}), (\ref{eq:temperature_z1}),
one can easily write the action for this system as\footnote{See more details in Appendix~\ref{app:gluon_action}.}
\begin{equation}
 S_E = \frac 1 {4 g_{YM}^2} \int d^3 x \int_0^{1/T(z_1)} d \tau_1  \left ( \frac 1  {1 +{\alpha(z_1)} (z-z_1)}  {\left({\bs E}^{a} \right)^2} + \left (1+{\alpha(z_1)} (z-z_1) \right ) \left({\bs H}^{a} \right)^2
 \right ) \,.
 \label{eq:SE_continuum_z1}
\end{equation}
Let us further assume that the acceleration $\alpha(z_1)$ is weak, i.e. the observer at $z=z_1$  is sufficiently far from the horizon. If we consider the volume of the size $\sim$few$\times \zeta$ in the vicinity of the observer at $z=z_1$, the factor $1+\alpha(z_1)(z-z_1)$  varies very slowly in this volume and  to the first approximation it  can be substituted by unity.  The only parameter that reminds us about the Rindler spacetime is the local temperature $T(z_1)$. One can expect that the size of $\sim$few$\times \zeta$ in the vicinity of the observer is sufficient for local thermalization in this volume.
So, we conclude that locally the Rindler spacetime is reduced to the gluodynamics in the flat space with the temperature depending on the $z$-coordinate. Notice that one can apply the same reasoning to the observer located at any point where the acceleration is weak. 
So, the system under consideration behaves like the standard gluodynamics with local properties determined by slowly varying local temperature. We call this property local thermalization. Notice that  similar phenomenon was observed in slowly rotating gluodynamics. However, in the case of rotating system the local action has more complex form, which can not be reduced to the change in local temperature~\cite{Braguta:2024zpi}. 

The idea of local thermalization for weak acceleration allows one to anticipate an interesting phenomenon  in gluodynamics in the Rindler space that can be explained as follows. Let us consider two observers. The temperature at the position of the first observer is 
$T_1>T_{c0}$ while the temperature at the position of the second one is $T_2<T_{c0}$, where $T_{c0}$ is the critical temperature of the confinement-deconfinement phase transition in the standard gluodynamics. So, the first observer is in the deconfinement phase, whereas the second one is in the confinement phase. In addition, it is reasonable to assume that the boundary between two phases is determined by the requirement that the local temperature (\ref{eq:temperature_z1}) is equal to $T_{c0}$. Thus one part of the volume resides in the deconfinement  spatially separated from the other part in the confinement. We call this phenomenon spatial confinement-deconfinement transition in gluodynamics. 

It should be noted here that the properties described above do not work close to the horizon, where the acceleration is sufficiently large $\alpha \sim \Lambda$.  For such a large acceleration the temperature is determined by the formula (\ref{eq:temperature_z1}) but one cannot ignore the variation of the factor  $1+\alpha(z_1)(z-z_1)$ in the action (\ref{eq:SE_continuum_z1}) and local thermalization of the system does not take place. 

We conclude this section by noting that the first study of accelerated gluodynamics was performed in Ref.~\cite{Chernodub:2024wis}. Rather than formulating the theory in Rindler spacetime, the authors considered gluodynamics with a position-dependent temperature $T(z)$, following the Tolman-Ehrenfest law~(\ref{eq:temperature_z1}).
While we demonstrate in Appendix~\ref{app:gluon_action} that this method is equivalent to the Rindler spacetime formulation for scalar field,  for vector gluon field  both methods are not fully compatible.
In particular, for gluodynamics, the approach of Ref.~\cite{Chernodub:2024wis} differs from that applied in this work by the corrections of order of $\sim \alpha/\Lambda$ or $\sim \alpha^2/\Lambda^2$  depending on observable.

\subsection{Lattice setup}\label{sec:Theory_setup}
To discretize the continuum action (\ref{eq:SE_continuum}) we are going to use the lattice approximation based on tree-level improved Symanzik action~\cite{Curci:1983an, Luscher:1985zq}.  In order to account the Rindler metric~\eqref{eq:rindler} in the gluon action we multiply the chromoelectric and chromomagtetic lattice plaquettes/rectangles by the factors $1/(1+\alpha_l z)$ and  $(1+\alpha_l z)$, respectively. Thus the lattice version of the action~(\ref{eq:SE_continuum}) reads 
\begin{multline}\label{eq:SE_lattice}
S_{E} = \beta  \sum_{x}\bigg[
\frac{1}{1 + \alpha_l z} \sum_{i} \bigg(c_0  \left(1 - \frac{1}{N_c} \text{Re} \Tr \bar{U}_{0i}\right) + 
c_1 \left(2 - \frac{1}{N_c} \text{Re} \Tr \big(\bar{W}_{0i}^{1\times2} + \bar{W}_{i0}^{1\times2}\big) \right) \bigg) + {}
 \\
{} +
(1 + \alpha_l z) \sum_{j>k} \bigg(c_0  \left(1 - \frac{1}{N_c} \text{Re} \Tr \bar{U}_{jk}\right) + 
c_1 \left(2 - \frac{1}{N_c} \text{Re} \Tr \big(\bar{W}_{jk}^{1\times2} + \bar{W}_{kj}^{1\times2}\big) \right) \bigg)
\bigg]\, ,
\end{multline}
where $\beta = 6/g_{YM}^2$ is an inverse lattice gauge coupling,  $\alpha_l=\alpha a$ is the acceleration in lattice units, $\bar{U}_{\mu\nu}$ denotes the clover-type average of four plaquettes, $\bar{W}_{\mu\nu}^{1\times2}$ is the clover-type average of four rectangular loops, and $c_0 = 1 - 8 c_1$, $c_1 = -1/12$. 

A key feature of the lattice action~\eqref{eq:SE_lattice} is that it does not suffer from the sign problem. Thus, one can apply lattice methods directly to study gluodynamics in the Rindler spacetime for real acceleration. 

We perform simulations on the lattices of the size $N_t \times N_x \times N_y \times N_z = N_t \times N_s^2 \times N_z$ (where $N_x = N_y = N_s$).
The lattice sizes used in our simulations are the following. We chose a base system of the size $N_t \times N_s^2 \times N_z = 5 \times 40^2 \times 121$. For continuum limit exploration, we calculate results for three more temporal lattice extensions, $N_t = 4, 6, 8$, keeping the ratio of $N_s, N_z$ to $N_t$ constant as $N_s/N_t = 8$ and $(N_z-1)/ N_t = 24$, respectively. 
To study the effects of finite volume, we vary the spatial lattice sizes $N_s$, $N_z$ independently, thus considering lattices with $N_s = 30, 50, 60$ and $N_z = 81, 151, 181$ in addition to our base lattice.
Unless otherwise specified, the results discussed in subsequent sections pertain to our base lattice.
To perform the Monte Carlo simulations, we use a heatbath algorithm supplemented by a few overrelaxation updates and accumulate statistics ranging from 100,000 to 400,000 sweeps, depending on the lattice parameters.

As was noted above in our study we consider the case of weak accelerations. In particular, the numerical values of the accelerations used in our simulations are $\alpha = 1, 2, 3, 6, 9, 18$~MeV. So, the inequalities {$\alpha\ll \Lambda$} or $\alpha \zeta \ll 1$ are always satisfied in our analysis.

The periodic boundary conditions are implemented in $\tau, x, y$-directions. 
Along the direction of acceleration, $z$, we impose open boundary conditions. The influence of boundary conditions on our results can be anticipated. However, we believe that disturbing effects coming from our boundary conditions   will be considerably suppressed because of screening for sufficiently large lattice size in the $z$-direction and  weak acceleration, which is fulfilled in our analysis. We are going to check this  below.
Notice that this statement was reliably confirmed in the investigation of rotating gluodynamics carried put in papers~\cite{Braguta:2021jgn, Braguta:2024zpi}. 
 
As discussed above, the action~(\ref{eq:SE_lattice}) assumes that the observer is located at $z=0$. We adjust the geometry of our lattices in such a way that the origin of the coordinate system is located at the center of our lattices in the $z$-direction. Thus, the $z$ coordinate in the physical units is varied in the region  $z \in [-(N_z - 1)a/2, (N_z - 1)a/2]$.\footnote{In our simulations $N_z$ is an odd number}  What concerns the other lattice directions, $x,y,\tau$, we implement usual coordinates for lattice simulations. 

For the geometry employed in this paper the horizon in the Rindler spacetime is located at $z_h=-1/\alpha$, i.e. it is remote from the position of the observer for weak accelerations. Moreover, our lattice parameters are adjusted in such a way that the whole lattice volume is distant from the horizon, and large accelerations $\alpha(z) \sim \Lambda$ (according to Eq.~(\ref{eq:acceleration_z1})) are never reached in studied volume.\footnote{Since the lattice volume under investigation is located far from the horizon and  acceleration is weak, one can expect that  quantum effects in gravity are either absent or considerably suppressed in our lattice setup. In particular, the Unruh temperature for our largest acceleration is $T_U = \alpha/2\pi \simeq  2.9$~MeV, which is much smaller than any temperature studied in our paper.} 

According to the Tolman-Ehrenfest law~\cite{Tolman:1930ona}, the thermal equilibrium temperature in gravitational field depends on the coordinate and for the Rindler spacetime this dependence is given by the formula~(\ref{eq:temperature_z1}). In this paper, we denote $T$ as the temperature at the position of the observer, $T=T_0=T(z=0)$, and it is equal to the inverse size of our lattices in the temporal direction $T=1/a N_t$. 

At the end of this section we would like to discuss the scale setting of lattice gluodynamics in the Rindler spacetime. In order to do this let us imagine 3+1 dimensional lattice which has fixed sizes in the $\tau, x, y$-directions and extends from the observer at $z=0$ to a very large $z$-coordinate. We also assume that for this setup the parameters are the lattice spacing  $a$, the inverse coupling constant  $\beta$, and  the acceleration $\alpha$ evaluated at $z=0$. The scale setting in the inhomogeneous gravitational background is nontrivial.  However, let us pass to the distant observer with sufficiently large $z$-coordinate and apply the property (\ref{eq:acceleration_z1}), (\ref{eq:temperature_z1}).  In this case we are left in the Rindler space with very small acceleration $\alpha(z)$. So, enhancing $z$-coordinate, one can make the Rindler metric arbitrary close to the Euclidean one. At the same time one has the same lattice and the parameters $a$ and $\beta$. For the distant observer and sufficiently large volume we can conduct the procedure of scale setting for the standard gluodynamics and determine  $\beta(a)$ (see Appendix~\ref{app:gluon_action}).  Thus, in our study of the gluodynamics in the Rindler spacetime we can use the scale setting from standard homogeneous gluodynamics. Notice that in our reasoning the lattice action is assumed to be sufficiently close to the continuum limit. In practice,  we take the values for string tension as a function of $\beta$ from Ref.~\cite{Beinlich:1997ia} with $\sqrt{\sigma} = 440$~MeV.
For reference, we will use the critical temperature $T_{c0}$ of the standard gluodynamics. We set this value using the infinite-volume extrapolation results for the critical coupling $\beta_c$ from Refs.~\cite{Beinlich:1997ia, Borsanyi:2022xml}.\footnote{To be more precise, we use the following values: $\beta_c({N_t = 4}) = 4.0730$, $\beta_c({N_t = 5}) = 4.20126$, $\beta_c({N_t = 6}) = 4.31447$, $\beta_c({N_t = 8}) = 4.51048$. }

\subsection{Observables}\label{sec:Observables}
In the Minkowski space, the Yang-Mills theory has two phases: at low temperature, the system is in the confinement phase, whereas at high temperatures it goes to the deconfinement phase.
To distinguish the phases the Polyakov loop is commonly used,
\begin{align} \label{eq:L_continuum}
    L({\bs r}) = \Tr\, \mathcal{P} \exp \biggl(\oint_{0}^{1/T} d \tau A_4( {\tau},{\bs r}) \biggr) \,,
\end{align}
where ${\mathcal P}$ is the path-ordering operator. The Polyakov loop is related to the free energy of static quark in gluon medium, $F_q$, as $L=e^{-F_q/T}$. In the confinement phase, the energy of a quark is infinite $F_q=\infty$, i.e. $L=0$, while in the deconfinement phase it is finite and $L\neq 0$. The confinement-deconfinement transition is related to the center $\mathbb{Z}_3$-symmetry of gluon action. In the deconfinement phase this symmetry is spontaneously broken, whereas in the confinement it persists. In $\textrm{SU}(3)$ gluodynamics, the confinement-deconfinement phase transition is of the first order. 

On the lattice the Polyakov loop can be calculated as follows,
\begin{equation}\label{eq:L_lattice_bare}
    L^b(\bs r) = \frac{1}{N_c} \Tr \left[ \prod_{\tau = 0}^{N_t - 1} U_4(\tau, \bs r) \right]\,,
\end{equation}
where $U_4(\tau, \bs r)$ is the link variable in the Euclidean time direction.
The expression~(\ref{eq:L_lattice_bare}) contains ultraviolet divergencies and need to be renormalized. In standard gluodynamics, the renormalization prescription of this operator is well elaborated (see, for instance, Ref.~\cite{Kaczmarek:2002mc}). Specifically the bare Polyakov loop, $L^b$, is multiplicatively renormalized as,
\begin{equation}\label{eq:L_ren}
    L = \left(Z(g^2)\right)^{N_t} L^b\,,
\end{equation}
where $Z(g^2)$ is the renormalization factor.

To study the phase transition in gluodynamics in the Rindler spacetime we are going to calculate first the bare Polyakov loop as a function of the $z$-coordinate,
\begin{equation}\label{eq:L_z_lattice}
    L^b(z) = \frac{1}{N_s^2 \delta z} \sum_{\bar z} \sum_{x,y} L^b(x,y,\bar z)\,,
\end{equation}
where $L(x, y, z) = L(\bs r)$.
Notice that in the last equation, in addition to the transverse coordinates $x$ and $y$ (perpendicular to the acceleration), we take an average over a thin layer along the direction of acceleration. To be more precise,  we average the Polyakov loop over the layer of thickness $\delta z_{\rm ph}  = \delta z \cdot a$ and  $z$ is the center of this averaged volume. 
In the calculations we use $\delta z = 1,\, \dots, 2 N_t$. Unless otherwise stated, results are presented for $\delta z =1$.

In Appendix~\ref{app:Renormalization}, we argue that instead of the equation~\eqref{eq:L_ren}, in the Rindler spacetime the following renormalization  of the local Polyakov loop can be used,
\begin{equation}\label{eq:L_z_ren}
    L(z) = \left(Z(g^2)\right)^{N_t (1+\alpha z)} L^b(z)\,.
\end{equation}
The values of the renormalization factor $Z(g^2)$ are taken from Ref.~\cite{Gupta:2007ax}.
Notice that, on the lattice, the modulus of the Polyakov loop, $\langle |L(z)| \rangle$,  is commonly used instead of $\langle L(z) \rangle$. Both observables coincide in the infinite volume limit. 
In finite volume, $\langle |L(z)| \rangle$ exhibits the confinement-deconfinement transition as a crossover from small, near-zero values to large values that deviate substantially from zero. In this paper, we are going to use $\langle |L(z)| \rangle$.

In order to detect the position of the spatial confinement-deconfinement transition, which will be referred to as the critical distance $z_c$, one can use the inflection point. To determine $z_c$, we fit the local Polyakov loop around the transition point with the function
\begin{equation}\label{eq:fit_tanh}
    L(z) = C_0 \tanh\left( \frac{z-z_c}{\Gamma}\right) + C_1\,,
\end{equation}
where $C_0$, $C_1$, $\Gamma$ and $z_c$ are the fit parameters.
Conducting the fitting, we take into account the correlations between adjacent points by estimating the covariance matrix using the jackknife method.
We also checked another function, $C_0 \arctan((z - z_c)/\Gamma) + C_1$, to calculate the transition point  and found that the difference in the critical distance $z_c$ between these two fitting functions is much  smaller then the statistical uncertainties.
So, we use the function~\eqref{eq:fit_tanh}, because it allows us to consider larger fitting regions with appropriate fit quality.

The properties of the spatial confinement-deconfinement phase transition can be investigated using the Polyakov loop susceptibility $\chi$. Similarly to the Polyakov loop, the local susceptibility depends on $z$-coordinate and can be written as 
\begin{gather}\label{eq:chi_z_ren}
T^3 \chi(z) =  \frac{N_s^2\, \delta z}{N_t^3} \left( \langle |L(z)|^2 \rangle  - \langle |L(z)|\rangle^2 \right)\,.
\end{gather}

In addition to the inflection point, the critical distance $z_c$ can be determined from the peak position of  $\chi(z)$.
Below we fit the local  susceptibility~\eqref{eq:chi_z_ren} around its peak with a Gaussian function,
\begin{equation}\label{eq:fit_Gauss}
    \chi(z) = A_0 \exp\left(-\frac{(z-z_c)^2}{2 \sigma^2}\right) + A_1 + A_2 (z-z_c)\,,
\end{equation}
and determine the peak position, $z_c$, width, $\sigma$ and height, $\chi^{\rm max}$, of the transition using the best fit parameters.
We consider the Gaussian function~\eqref{eq:fit_Gauss} with additional linear slope, $A_2 \neq 0 $, because it better describes the data and allows us to use more points around the peak.
To estimate a systematic uncertainty due to the specific choice of the fitting function, we also use the function~\eqref{eq:fit_Gauss} with $A_2 = 0$, in a range where it is applicable.
As a final result, we use a $p$-value based average of transition parameters from two fitting functions and different ranges of data points involved in the fitting procedure.
The uncertainties of the data presented below include both statistic and systematic contributions.

The scaling properties of the width and the height of the susceptibility peak can be used to determine the type of the phase transition.
Notice, however, that in homogeneous statistical system the width of the transition is defined along the temperature axis.
In the Rindler spacetime, $\sigma$ is the width of spatial confinement-deconfinement transition in coordinate space, which assumes another interpretation of the results.
This properties of the spatial transition will be discussed in Section~\ref{sec:Results_Crossover} in the detail.

\section{Results}\label{sec:Results}

This section presents the results and analysis from our lattice simulation for accelerated gluodynamics. 
In particular, we focus on studying of the confinement-deconfinement phase transition in the Rindler space.
It is important to notice that all physical quantities  are considered  from the perspective of the observer located at $z=0$ coordinate.

\subsection{Spatial confinement-deconfinement transition and Tolman-Ehrenfest law}
\label{sec:Results_TE_checking}

\begin{figure*}[t]
    \centering
    \includegraphics[width=0.49\linewidth]{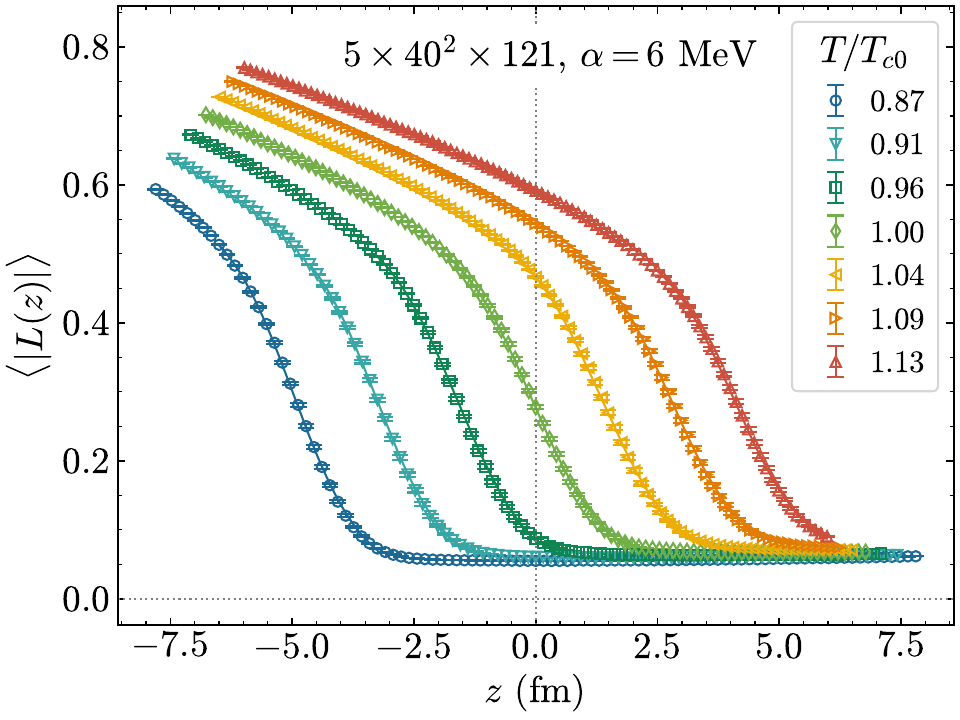}
    \hfill
    \includegraphics[width=0.49\linewidth]{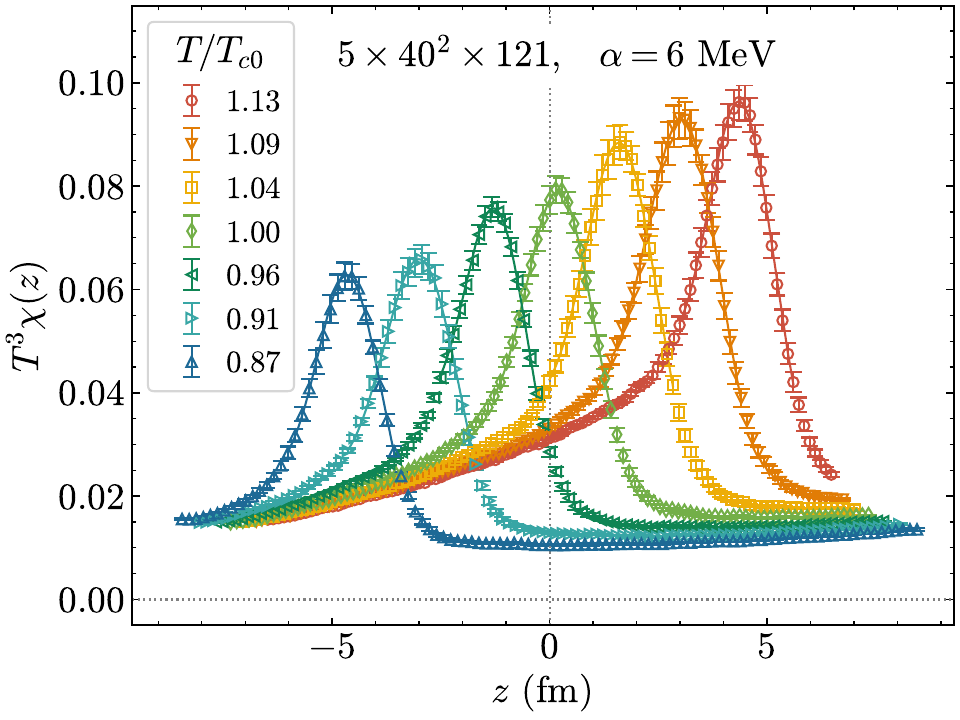}
    \caption{%
    The renormalized local Polyakov loop (left) and its susceptibility (right) as a function of coordinate $z$ for different temperatures.
    The results were calculated on the lattice $5\times 40^2\times 121$ for $\alpha = 6$~MeV.
    }
    \label{fig:L_vs_z_nt5_a6MeV}
\end{figure*}

Figure~\ref{fig:L_vs_z_nt5_a6MeV} shows the modulus of the renormalized local Polyakov loop~\eqref{eq:L_z_ren} and the susceptibility~\eqref{eq:chi_z_ren} as a function of the spatial coordinate~$z$ along the direction of acceleration.\footnote{Several lattice points adjacent to the boundary were discarded because they are affected by open boundary conditions.}
The results are presented for seven different temperatures around the critical one $T_{c0}$ of the standard gluodynamics with the acceleration fixed at $\alpha=6$~MeV on the base lattice $5\times 40^2 \times 121$.
From Fig.~\ref{fig:L_vs_z_nt5_a6MeV}, one can see that the Polyakov loop is close to zero in the right region of the figure, which can be associated with confinement phase, while $\langle |L(z)| \rangle$ deviates significantly from zero in the left region which can be associated with deconfinement phase.
Thus we observe coexisting phases and spatial confinement-deconfinement  transition for all temperatures under consideration.
In particular, for the temperature equals to the critical one of the standard gluodynamics, $T = T_{c0}$, the transition occurs approximately at the observer's position, $z = 0$. 
The region $z \lesssim 0$, which is closer to the horizon, is in the deconfinement phase, while the region $z \gtrsim 0$ is in the confinement phase.
The transition point is shifted to the region $z>0$ for $T>T_{c0}$ and to region $z<0$ for $T<T_{c0}$.

Qualitatively, this physical picture is in agreement with the Tolman--Ehrenfest law~\eqref{eq:temperature_z1}.
Indeed, according to the formula~(\ref{eq:temperature_z1}), the regions with smaller $z$-coordinates are more heated as compared to the regions with larger $z$-coordinates. As a result, for a certain parameters $T$ and $\alpha$, the left part of the system under consideration passes to the deconfinement whereas the right part remains in the confinement phase. 
To describe this behavior quantitatively it is reasonable to assume that the transition between phases takes place at the coordinate $z_c^{\rm TE}$ which can be found from the equation  $T(z_c^{\rm TE}) = T_c(\alpha)$. 
In last equation we assumed that the critical temperature $T_c(\alpha)$ may itself depend on acceleration, but for small accelerations this dependence can be disregarded, and to the first approximation one can take $T_c(\alpha) \approx T_{c0}$.
This condition leads to the expression
\begin{equation}\label{eq:zc_TE}
    z_c^{\rm TE} = \frac{1}{\alpha}\left( \frac{T}{T_{c0}} -1 \right)\,.
\end{equation}

On the lattice we determine the critical distance $z_c$, i.e. the boundary between  confinement and deconfinement phases, from the inflection point of the local Polyakov loop (see Section~\ref{sec:Observables}). In addition to the inflection point we utilized the susceptibility of the Polyakov loop \eqref{eq:chi_z_ren} for which the critical distance is determined from the susceptibility peak. The results obtained in this way are in good agreement with that extracted from the inflection point (see the detailed comparison in Appendix~\ref{app:chi_vs_L_vs_bare}). In this section we will focus on the analysis of the critical distance calculated from the local Polyakov loop with $\delta z = 1$.
We found that our results on the critical distance weakly depend on $\delta z$ in the studied interval $[1,2 N_t]$.

\begin{figure}[t]
    \centering
    \includegraphics[width=0.49\linewidth]{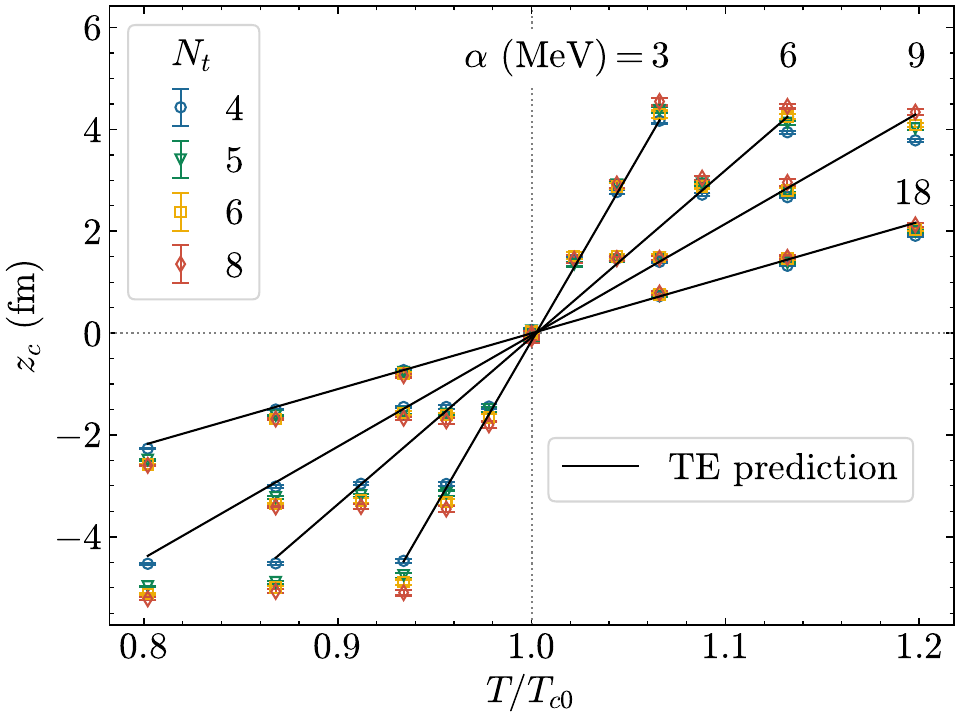}
    \caption{%
    The critical distance $z_c$ as a function of temperature for various accelerations $\alpha$. These results were obtained on the lattices $4\times 32^2 \times 97$, $5\times 40^2 \times 121$, $6\times 48^2 \times 145$, $8\times 64^2 \times 193$. The solid black lines represent the TE-prediction~\eqref{eq:zc_TE}.
    }
    \label{fig:zc_vs_T}
\end{figure}

The critical distance $z_c$ as a function of  the ratio~$T/T_{c0}$ is shown in Fig.~\ref{fig:zc_vs_T}. The data correspond to lattices with fixed aspect ratio $N_s/N_t = 8$ and various temporal sizes $N_t = 4, 5, 6, 8$, for the accelerations $\alpha = 3, 6, 9, 18$~MeV. The black solid lines in Fig.~\ref{fig:zc_vs_T} show the TE-prediction (\ref{eq:zc_TE}). It is seen from this figure that  the lattice data with small deviations are in good agreement with the TE law. 

To study the deviation of the lattice data from the TE-prediction \eqref{eq:zc_TE}  we fitted the data by the formula
\begin{equation}
\label{eq:zc_fit}
    z_c = \frac{k_0}{\alpha} \left(\frac{T}{T_{c}}  - 1\right) + \frac{k_1}{\alpha} \left(\frac{T}{T_{c}}  - 1\right)^2\,,
\end{equation}
where $k_0$, $k_1$, and $T_c$ are fitting parameters.\footnote{The critical temperature of standard gluodynamics $T_{c0}$  in the formula \eqref{eq:zc_TE} is not a fitting parameter. We took $T_{c0}$ from papers \cite{Beinlich:1997ia, Borsanyi:2022xml}. On the contrary, $T_c$ in Eq.~\eqref{eq:zc_fit} is the fitting parameter which accounts possible influence of the acceleration and finite volume effects on the critical temperature.}
The function \eqref{eq:zc_fit} describes lattice data well and one can determine values of $k_0$, $k_1$ and $T_c$ for all lattice parameters used in the simulations.

\begin{figure*}[t]
    \centering
    \includegraphics[width=0.49\linewidth]{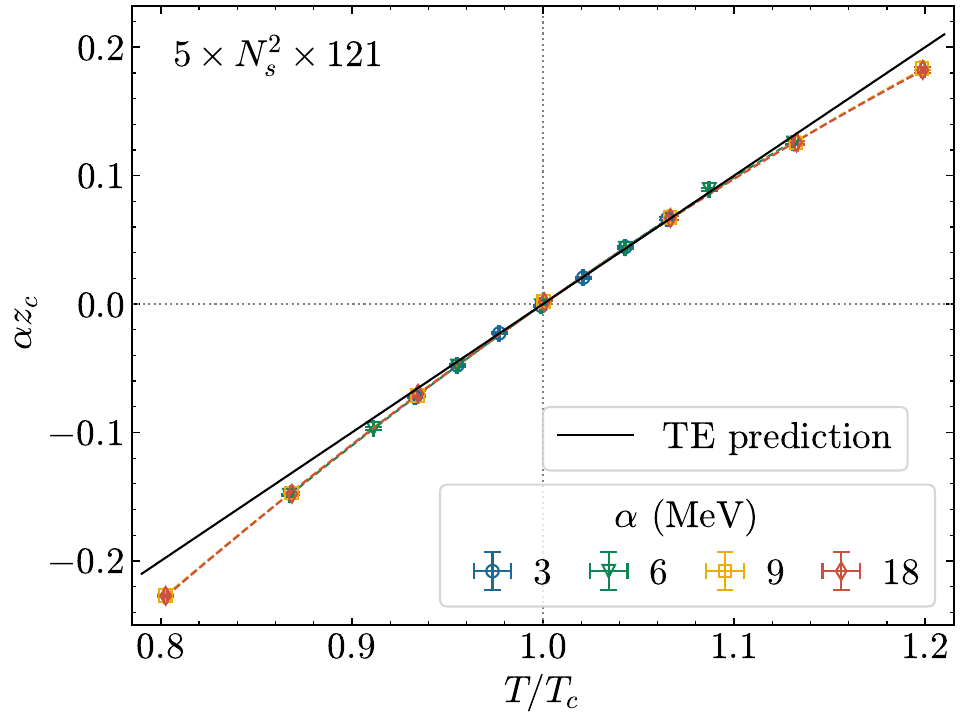}   
    \hfill
    \includegraphics[width=0.49\linewidth]{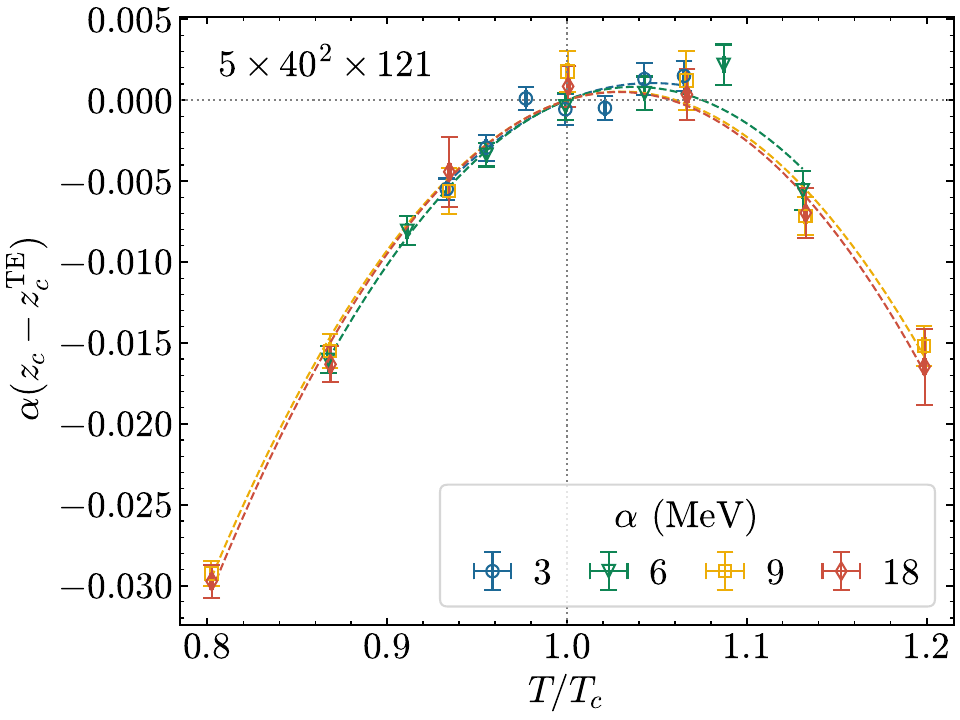}
    \caption{%
    (left) The critical distance normalized by acceleration, $\alpha z_c$, as a function of temperature for various accelerations $\alpha$. The results were obtained on the lattice $5\times 40^2 \times 121$. Dashed lines represent the fit of the data by quadratic function~\eqref{eq:zc_fit}, and black solid  line shows the TE-prediction, $z_c^{\rm TE}$ from Eq.~\eqref{eq:zc_TE}.
    (right) The difference between  $z_c$ and $z_c^{\text{TE}}$  as a function of temperature.
    The horizontal error bars represent the uncertainty in temperature, which propagates from the fit parameter $T_c$. 
    }
    \label{fig:zc_vs_T_scaled}
\end{figure*}

In Fig.~\ref{fig:zc_vs_T_scaled}~(left), we plot the product $\alpha z_c$ as a function of the ratio $T/T_c$ for various accelerations. The data shown on this figure are obtained on the lattice $5\times 40^2\times 121$. Notice that the value of  $T_c$, which was used for horizontal axis, was obtained from the fitting of the lattice data by the formula~\eqref{eq:zc_fit}. As a result, each data point contains small statistical uncertainty in the ratio $T/T_c$. It is seen from Fig.~\ref{fig:zc_vs_T_scaled}~(left), that after the rescaling of the critical distance: $\alpha z_c$, the data for all accelerations under consideration lie on one universal curve. 
The black solid line on Fig.~\ref{fig:zc_vs_T_scaled}~(left) represents the TE prediction, whereas the dashed lines shows the best fit of lattice data by the function~\eqref{eq:zc_fit}. One can see that TE law describes the lattice data quite well. However, small deviations from the TE law are noticeable for temperatures from $T_{c}$.

In order to study the deviation from the TE law, in Fig.~\ref{fig:zc_vs_T_scaled}~(right) we show the rescaled difference $\alpha (z_c - z_c^{TE})$ as a function of $T/T_c$. 
For each set of lattice parameters, the TE law prediction $z_c^{TE}$ was calculated according to the formula $\alpha z_c^{\rm TE} = \left(T/T_{c} - 1 \right)$, where the best fit value of  $T_c$ was used.
Interestingly, the deviation exhibits a parabolic dependence on temperature, which is captured by the parameters $k_0$ and $k_1$ in the equation~\eqref{eq:zc_fit}. It is seen that the characteristic size of the deviation is quite small.
The largest values of the difference $\alpha (z_c - z_c^{TE})$ is reached at the edges of the considered temperature interval, where the relative deviation is as large as $|(z_c - z_c^{\rm TE})/z_c^{\rm TE} | \sim 10-15\%$.

\begin{figure}[t]
    \centering
    \includegraphics[width=0.49\linewidth]{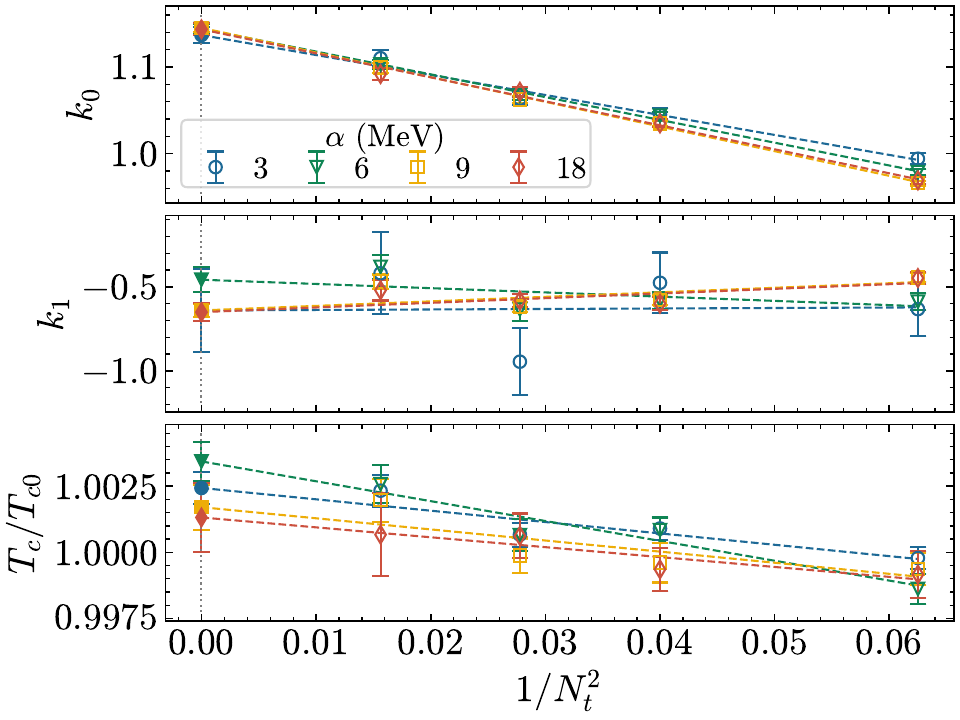}
    \caption{%
    The best fit parameters $k_0$, $k_1$ and $T_c$ as a function of $1/N_t^2$ for the accelerations under consideration. These results were obtained on the lattices  $4\times 32^2 \times 97$, $5\times 40^2 \times 121$, $6\times 48^2 \times 145$, $8\times 64^2 \times 193$. 
    Open markers show the parameters at particular $N_t$, dashed lines represent continuum-limit extrapolation, and filled markers show the values at the continuum limit.
    } 
    \label{fig:ki_Tc_coefficients}
\end{figure}

The best-fit parameters $k_0$, $k_1$ and $T_c$ for the studied lattices~--- with temporal extents $N_t = 4, 5, 6, 8$ and fixed aspect ratios $N_s/N_t = 8$ and $(N_z-1)/N_t = 24$~--- are presented in Fig.~\ref{fig:ki_Tc_coefficients}. The figure also shows the continuum limit extrapolation  of these parameters, obtained using the ansatz $\mathcal{O} = \mathcal{O}_0 + \mathcal{O}_1 / N_t^2$. 
In addition, we investigated the finite volume effects associated with transversal and longitudinal lattice sizes $N_s$ and $N_z$. These results are presented in Appendix~\ref{app:finite_volume} and they indicate weak dependence of our results on the $N_s$ , $N_z$ and boundary conditions.

For the studied accelerations, the leading coefficient~$k_0$ in the continuum limit is greater than unity  by $\sim 15\%$, which to be compared with the TE law prediction  $k_0=1$, Eq.~\eqref{eq:zc_TE}.
This disagreement might be attributed to the renormalization of the acceleration by the strong interaction. Another possible source of this disagreement might result from the fact that the confinement-deconfinement phase transition  becomes spatial crossover in the Rindler spacetime (see the next section). So, the pseudocritical temperature in such systems is not related to a singular point, and depends on the observable used for the calculation. 
The next quadratic coefficient~$k_1$ is negative and has larger uncertainty, which does not allow us to establish a clear trend with acceleration.
Notice, however, that the negative value of the $k_1$ coefficient partially compensates a bit excessive  value of  $k_0$.

The last parameter in the fitting formula~\eqref{eq:zc_fit} is  $T_c$ which accounts the finite volume effects and possible influence of the acceleration on the critical temperature.
From Fig.~\ref{fig:ki_Tc_coefficients} it is seen that in the continuum limit  $T_c$ is larger than the critical temperature of standard gluodynamics $T_{c0}$ by the value smaller than $0.3\%$. 
If one regards the uncertainty in these results and the fact  that the limit  $N_s \to \infty$ reduces the $T_{c}$ by approximately the same amount (see Fig.~\ref{fig:zc_vs_T_testNs}), one can draw a conclusion that we don't see the influence of the acceleration on the critical temperature.
This result is in accordance to the paper~\cite{Chernodub:2025ovo}, where the change of the critical temperature originating from the acceleration was estimated as $\delta T_c(\alpha)/T_{c0} \sim   (\alpha/ 2\pi T_{c0})^2/2 \sim 2\cdot10^{-5}$ for the acceleration $\alpha \sim 10$~MeV. 

At the end of this section it is interesting to compare spatial confinement-deconfinement  transition in rotating gluodynamics~\cite{Braguta:2023iyx, Braguta:2024zpi} with that observed in this paper. In Refs.~\cite{Braguta:2023iyx, Braguta:2024zpi} the authors carried out lattice study of gluon system in the rotating reference frame, where external gravitational field appears. 
Similarly to the Rindler spacetime, one can apply TE law in rotating system, which gives the local equilibrium temperature $T(r) \sim 1/\sqrt {1-\Omega^2 r^2}$, where $\Omega$ is the angular velocity of the rotating  frame and $r$ is the distance to the rotating axis.
Thus, in this case, the TE law predicts that rotation effectively heats up the system at increasing $r$.
Consequently, it could be anticipated that for rotating gluodynamics the phase arrangement looks as follows: confinement in the center and deconfinement at the periphery.
However, the lattice investigation show the opposite behavior, which contradicts to the TE law. An explanation, proposed and also confirmed in Ref.~\cite{Braguta:2024zpi}, suggests that gravitational field modifies the gluon action leading to the asymmetry between chromoelectric and chromomagnetic fields.
For this modified action the confinement phase appears in the periphery while deconfinement phase in the center of the studied system and TE law does not play a major role for rotating gluodynamics. 

One might assume that similar phenomenon might take place in the Rindler space time, since the Rindler metric gives rise to the asymmetry between chromoelectric and chromomagnetic fields in the action~\eqref{eq:SE_continuum}.
However, in the Rindler spacetime one can change the position of the observer and locally get rid of this asymmetry for weak accelerations (see Eqs.~\eqref{eq:acceleration_z1}-\eqref{eq:SE_continuum_z1} and Appendix~\ref{app:gluon_action}). For this reason, the asymmetry in the gluon action in the Rindler space does not play a significant role, and the TE law becomes the dominant. 

\subsection{Properties of the spatial confinement-deconfinement transition}
\label{sec:Results_Crossover}

It is known that confinement-deconfinement phase transition in the standard SU(3) gluodynamics is of the first order~\cite{Fukugita:1989yw}.  We have discovered above that in the Rindler spacetime the confinement-deconfinement becomes spatial transition. In this section we are going to investigate its properties. 

The first question to be addressed is the type of the spatial confinement-deconfinement transition. In this respect it is commonplace to study the scaling properties of the width $\sigma_T$, and height  $\chi^{\text{max}}_T$, of the susceptibility peak.
The subscript $T$ emphasize that in homogeneous statistical system the parameters of the transition are determined along the temperature axis.
In particular, for the first order phase transition their scaling properties  are $\sigma_T \sim 1/V$ and $\chi^{max}_T \sim V$~\cite{Binder:1984llk}. The second order phase transition manifests itself as $\sigma_T \sim 1/V^{-1/d\nu}$ and $\chi^{max}_T \sim V^{\gamma/d\nu}$ where $\gamma$ and $\nu$ are the critical exponents and $d$ is the dimension of the system~\cite{Svetitsky:1982gs,Engels:1995em}. In the case of crossover, both $\sigma_T$ and $\chi^{max}_T$ remain finite in the infinite volume limit. 

In the Rindler spacetime one has the width of spatial confinement-deconfinement transition in $z$-coordinate, $\sigma$, contrary to the width $\sigma_T$ along the temperature axis for homogeneous statistical system.  To acquire a better understanding of the relation between  $\sigma_T$ and $\sigma$,  let us consider the susceptibility $\chi(T)$ of the standard gluodynamics in finite volume, which  can be represented as 
\begin{equation} 
\label{eq:susceptibility}
\chi(T) = \chi^{max}_T \exp { \left ( - \frac {(T-T_{c0})^2} {2 \sigma_T^2} \right ) }\,, 
\end{equation}
in the vicinity of the critical temperature $T_{c0}$.
Here it is assumed that $\chi^{max}_T$ and  $\sigma_T$ depend on volume. 
According to the idea of local thermalization, described in Section~\ref{sec:Theory_Rindler}, for small accelerations, gluodynamics in the Rindler spacetime behaves like the standard gluodynamics in the volume $V \sim \mbox{few} \times \zeta (a N_s)^2$ at the local temperature determined by the TE law. Consequently, locally the formula \eqref{eq:susceptibility} is applicable. So it is reasonable to substitute the TE law \eqref{eq:temperature_z1} to the equation \eqref{eq:susceptibility} and get the susceptibility in the Rindler spacetime
\begin{equation}
\label{eq:susceptibility_rindler}
\chi(z) = \chi^{max} \exp { \left ( - \frac {z^2} {2 \sigma^2} \right ) }\,, 
\end{equation}
where 
\begin{equation}
\label{eq:width_rindler}
     \sigma = \frac {\sigma_T} {T_{c0}} \frac{1+\alpha z } {\alpha}\biggl |_{z \sim 0} \simeq \frac {\sigma_T} {T_{c0}} \frac{1} {\alpha}\,. 
\end{equation}
In the derivation of the equation \eqref{eq:susceptibility_rindler} it was assumed that we consider the vicinity of the observer's position and the temperature at  $z=0$ is adjusted to the critical one $T_{c0}$.  It is worth to mention that Eq.~(\ref{eq:susceptibility_rindler}) indicates that the peak of the susceptibility in temperature region of standard gluodynamics becomes a peak at the corresponding $z$-coordinate in the Rindler spacetime. 

It might seem from the derivation of Eq.~\eqref{eq:susceptibility_rindler} that the width $\sigma_T$ in Eq.~\eqref{eq:width_rindler} and height $\chi^{max}_T$ are the same as in the standard gluodynamics. Notice that these quantities depend on the volume. 
Since the equilibrium temperature depends on the coordinate $z$, it is unclear which size in the direction $z$ should be taken in the calculation of the effective volume $V$, and how it depends on the acceleration. Moreover, interactions in the Rindler spacetime might noticeably modify the scaling of standard gluodynamics. Thus one can draw a conclusion that $\sigma_T$ and  $\chi^{max}_T$ are not the same as in the standard gluodynamics. To mark this fact instead of $\chi^{max}_T$ in \eqref{eq:susceptibility_rindler}  new parameters  $\chi^{max}$ was introduced.
To summarize, in what follows we are going to use the formula \eqref{eq:susceptibility_rindler} with arbitrary parameters $\sigma$ and $\chi^{max}$ and determine them from fitting procedure.  

\begin{figure*}[t]
    \centering
    \includegraphics[width=0.49\linewidth]{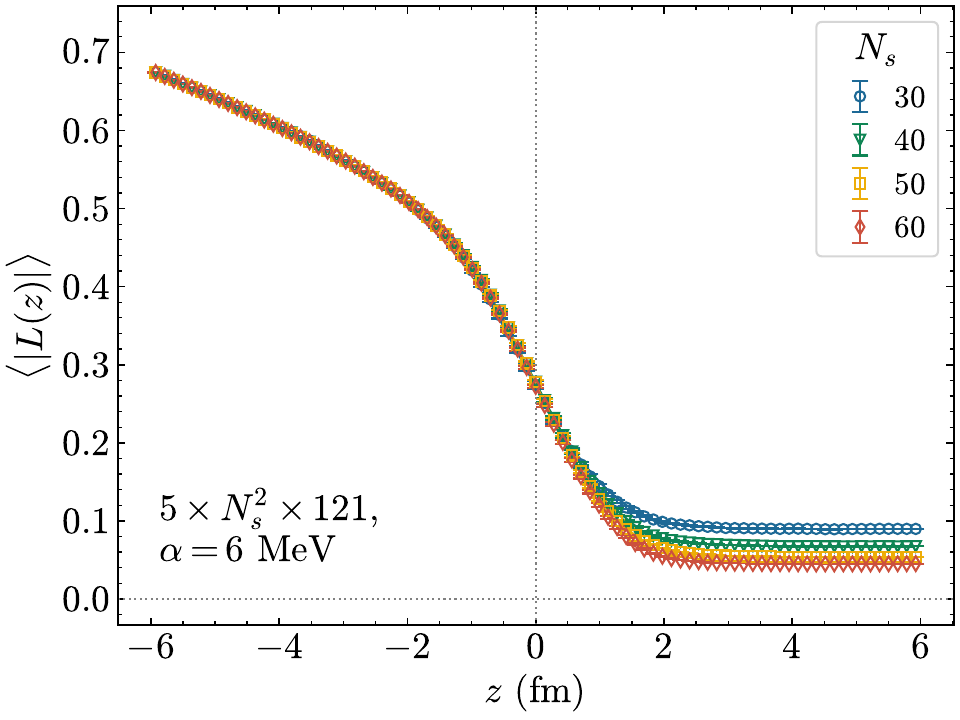}
    \hfill
    \includegraphics[width=0.49\linewidth]{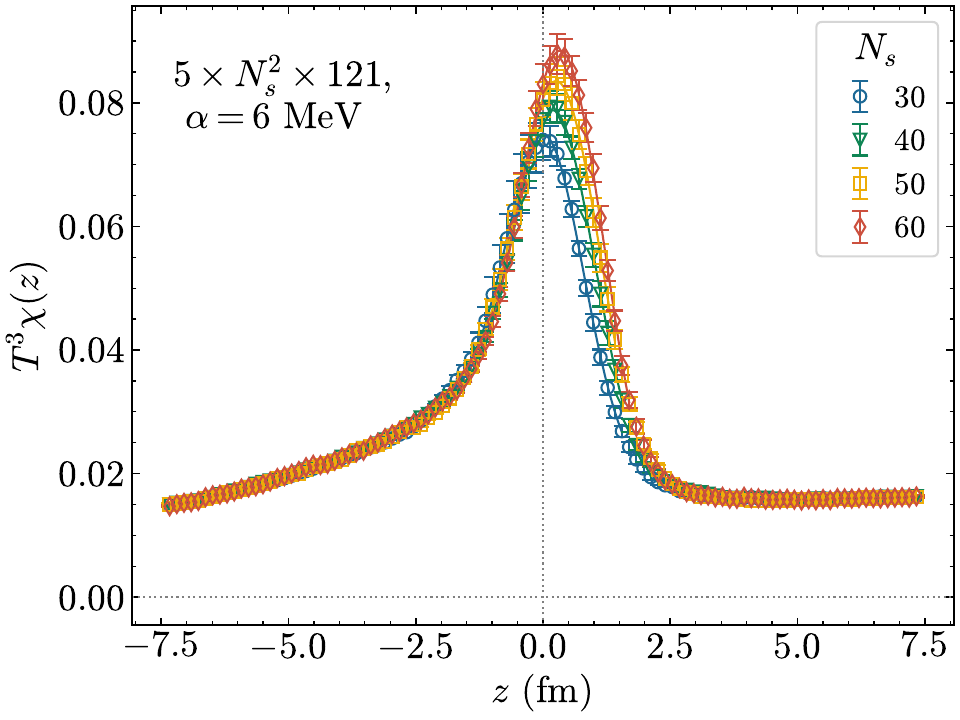}
    \caption{%
    The renormalized local Polyakov loop (left) and its susceptibility (right) as a function of coordinate $z$ for lattices $5\times N_s^2\times 121$ with various transverse sizes $N_s = 30, 40, 50, 60$ at acceleration $\alpha=6$~MeV and temperature $T = T_{c0}$.
    }
    \label{fig:Lz_vs_volume}
\end{figure*}

Now let us proceed to the lattice results. In order to study the type of the phase transition we adjusted the temperature at $z=0$ to the critical value of the standard gluodynamics $T=T_{c0}$.  In Fig.~\ref{fig:Lz_vs_volume}, we show the modulus of the local Polyakov loop (left panel)   and its susceptibility (right panel) as a functions of $z$ coordinate. These data were obtained on the lattices with fixed $N_t = 5$, $N_z = 121$ and various transverse sizes $N_s = 30, 40, 50, 60$ at $\alpha = 6$~MeV with $\delta z =1$.  
The local Polyakov loop demonstrates transverse volume dependence in the confinement phase which can be attributed to finite volume effects.
For the susceptibility, the volume dependence is visible only near the transition point.
However, the height and width of the peaks exhibit only a weak volume dependence, indicating the crossover nature of the transition. This conclusion is in agreement with the result of paper \cite{Chernodub:2024wis}. 

\begin{figure*}[t]
    \centering
    \includegraphics[width=0.49\linewidth]{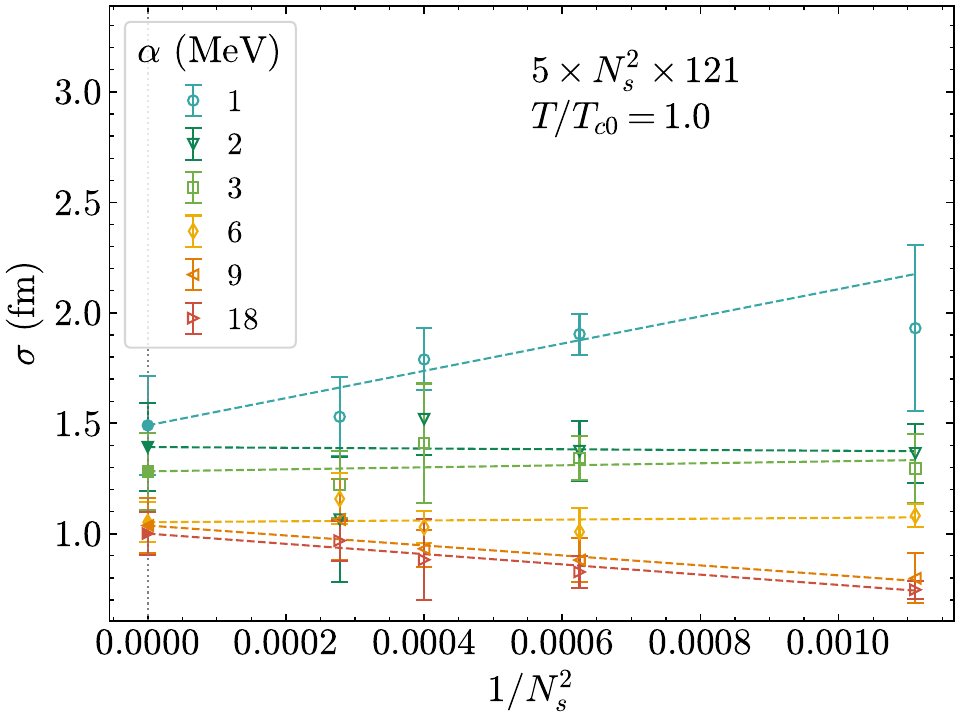}
    \hfill
    \includegraphics[width=0.49\linewidth]{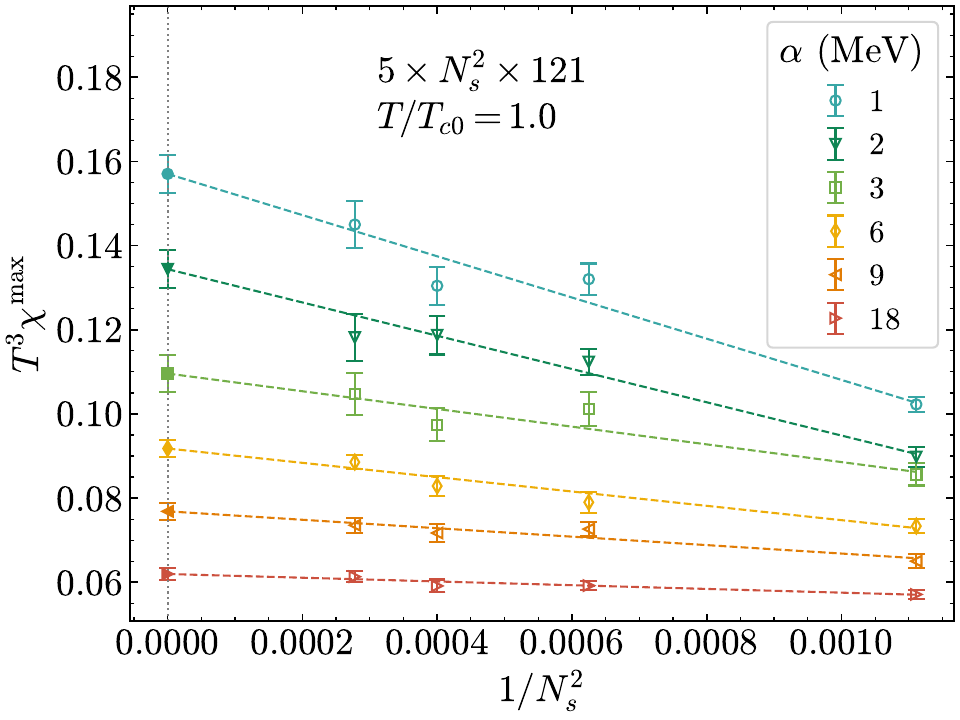}
    \caption{%
    The width $\sigma$ (left) and the height $\chi^{\text{max}}$ (right) of the susceptibility peaks as a function of $1/N_s^2$ at temperature $T = T_{c0}$. The results were obtained on the lattices $5\times N_s^2\times 121$ with $N_s = 30, 40, 50, 60$ at $\alpha = 1, 2, 3, 6, 9, 18$~MeV.
    }
    \label{fig:hw_vs_volume}
\end{figure*}

In Fig.~\ref{fig:hw_vs_volume}, we plot the width $\sigma$ (left panel), and height  $\chi^{\text{max}}$ (right panel) of the susceptibility as functions of inverse transverse volume, $1/N_s^2$, for various accelerations $\alpha$.
It is seen that the data are well described by a linear function of the form $\mathcal{O} = \mathcal{O}_0 + {\mathcal{O}_1}/{N_s^2}$, confirming that they remain finite in the infinite volume limit, $N_s\to \infty$, within the studied range of  accelerations. 
Note that a mild behavior of the fit parameters with the volume confirms that considered transverse size of the base lattice, $N_s = 40$, is sufficient for lattice simulation. 

Here it is worth to discuss the following issue. From Fig.~\ref{fig:hw_vs_volume} (left panel) one sees that  the width of the transition remains finite, when the infinite volume limit is taken through $N_s \to \infty$. Depending on the value of acceleration used in our study, this width is $\sim 1.0-1.5$ fm, which is  of order of  the correlation length in the standard gluodynamics. 
Based on this fact it is reasonable to assume that there exists the smallest width that separates  confinement from deconfinement phase in the Rindler spacetime and it is natural to expect that it is of order of the correlation length in gluodynamics. In this case the crossover nature of the spatial confinement-deconfinement transition can be explained by the existence of this smallest width. 
On the contrary, if one assumes that the spatial transition is of the first or the second order, then the width of the transition vanishes in the infinite volume limit $N_s \to \infty$. Physically this implies that there is a two dimensional plane with confinement phase on the one side while deconfinement on the other side. If one puts a static color charge to the deconfinement phase in the vicinity of this plane, gluodynamics must provide the screening of this color charge in the confinement phase within an infinitely small distances in the limit $N_s \to \infty$.
The physical pattern where the screening is determined by infinite volume scaling rather than by the dynamics of the theory seems unnatural. 
For this reason, we believe that the idea of finite transition width is more physically motivated and it is supported by our lattice results.  Notice that spatial confinement-deconfinement transition in rotating gluodynamics was also found  to be a crossover~\cite{Braguta:2023iyx, Braguta:2024zpi}. 
We believe that this fact is also related to the existence of the smallest transition width in gluodynamics.

\begin{figure*}[t]
    \centering
    \includegraphics[width=0.49\linewidth]{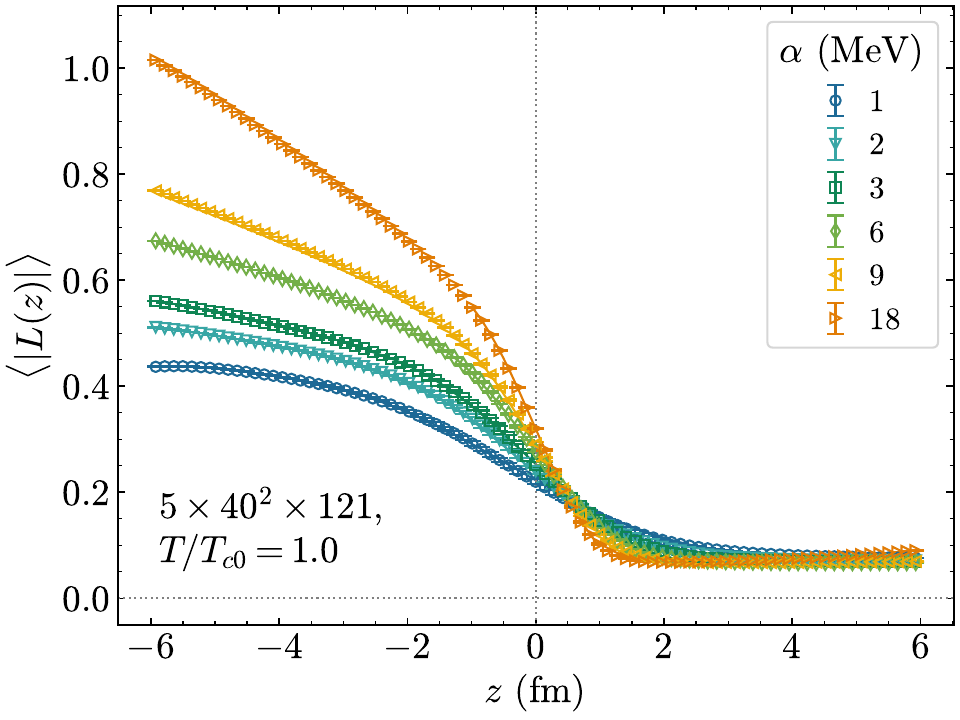}
    \hfill
    \includegraphics[width=0.49\linewidth]{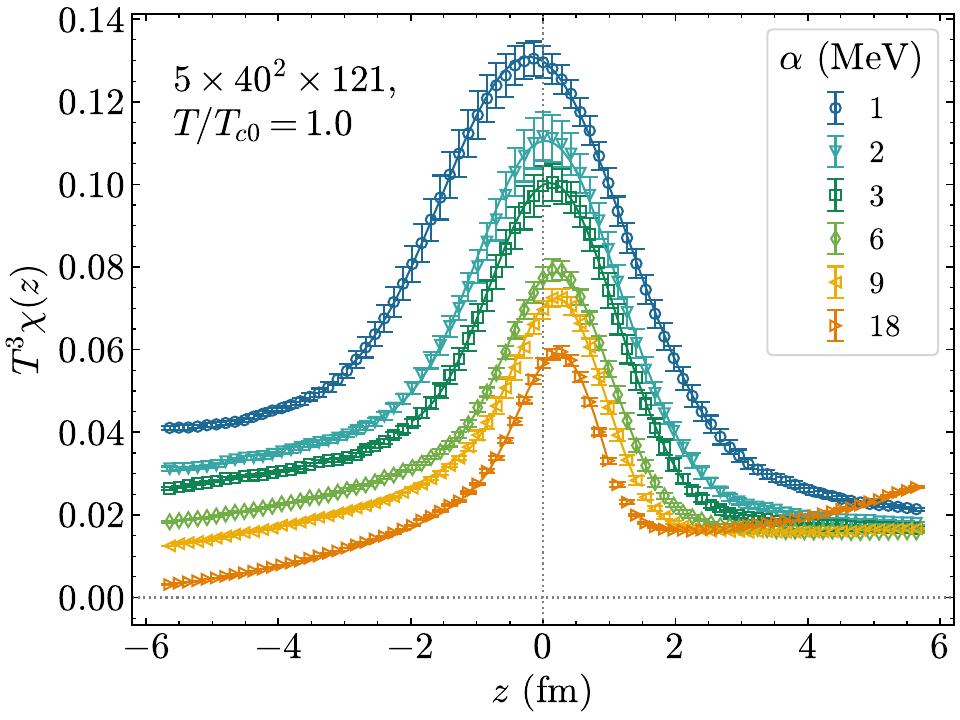}
    \caption{%
    The renormalized local Polyakov loop (left) and its susceptibility (right) as a function of coordinate $z$ for lattice $5\times 40^2\times 121$, for various accelerations $\alpha = 1,\,2,\,3,\,6,\,9,\,18$~MeV at temperature $T/T_{c0} = 1$.
    }
    \label{fig:Lz_vs_a}
\end{figure*}

\begin{figure*}[t]
    \centering
    \includegraphics[width=0.49\linewidth]{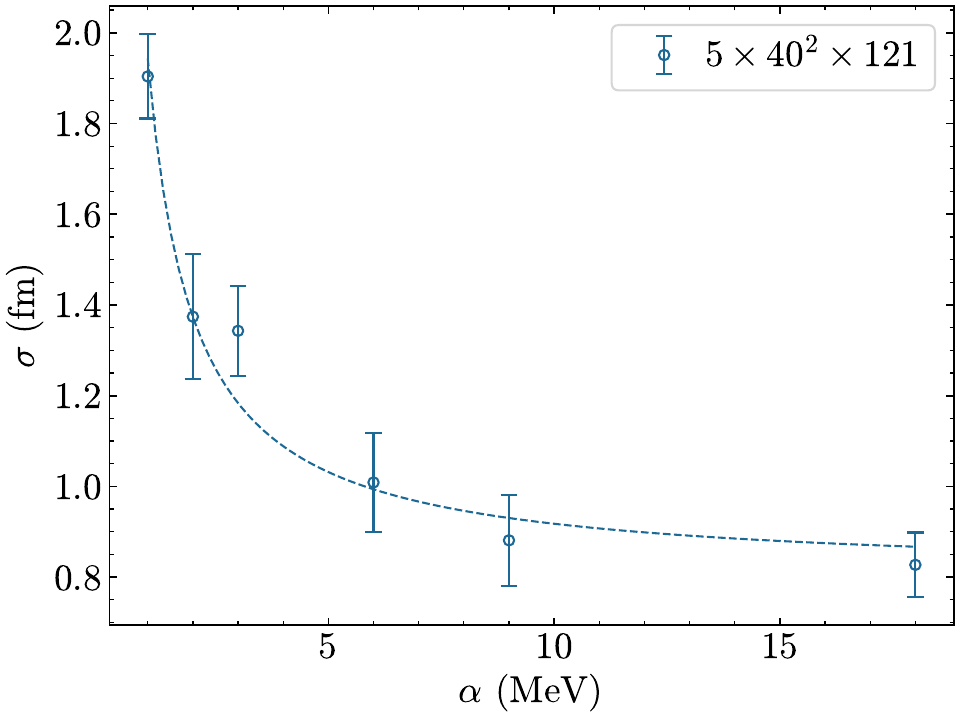}
    \hfill
    \includegraphics[width=0.49\linewidth]{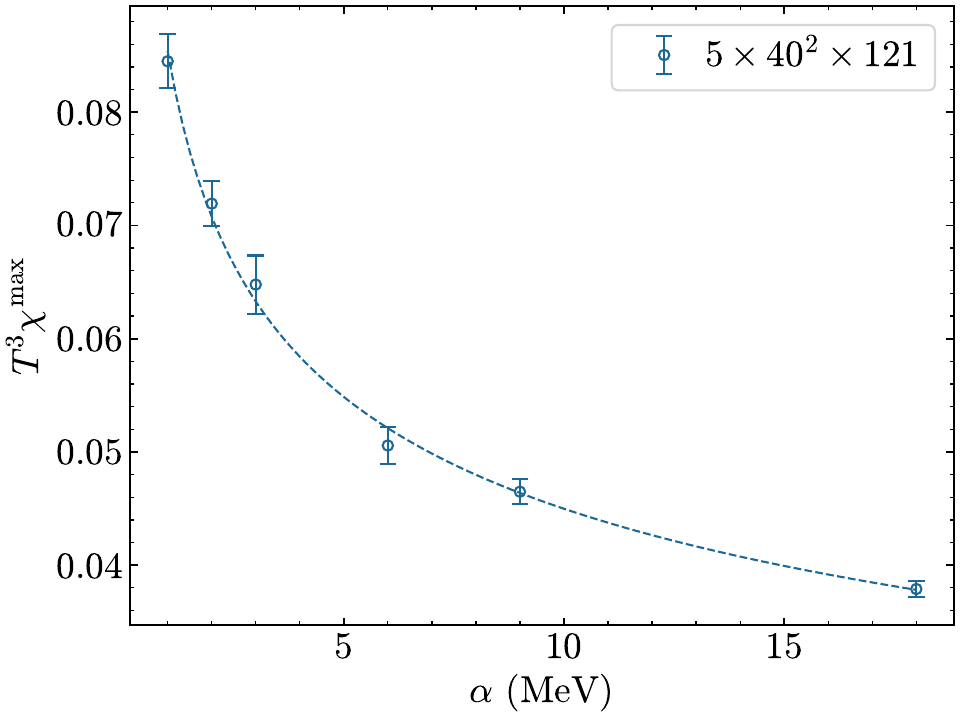}
    \caption{%
    The width $\sigma$ (left) and height $\chi^\text{max}$ (right) of the susceptibility peaks at temperature $T/T_{c0} = 1$ as a function of acceleration. Data are shown as open circles, and fits are shown as dashed lines.
    }
    \label{fig:hw_vs_a}
\end{figure*}

To study how the properties of the spatial transition change with  acceleration, in Fig.~\ref{fig:Lz_vs_a} we show the modulus of local Polyakov loop (left panel) and the susceptibility (right panel) as a function of $z$. The results are calculated  on the lattice $5\times 40^2 \times 121$ for various accelerations under consideration. From the data shown in Fig.~\ref{fig:Lz_vs_a}, we determine the susceptibility width and height for various accelerations at temperature $T=T_{c0}$. 
These quantities, $\sigma$ and  $\chi^{max}$, are presented in Fig.~\ref{fig:hw_vs_a}  as functions of acceleration. 
We fit these data by the functions 
\begin{gather}
\label{eq:sigma}
\sigma = b_0 + b_1 \frac {T_{c0}} {\alpha}\,,  \\
\label{eq:chi}
T^3\chi^{max} = d_0 + d_1 \left ( \frac {T_{c0}} {\alpha} \right )^{d_2}\,.
\end{gather}
The data are well described by these functions ($\chi^2/\mbox{ndf} \sim 1$) and the best fit parameters are $b_0=0.80(5)$~fm, $b_1=0.0041(4)$~fm, and $d_0=-0.009(24),~ d_1=0.02(2),~ d_2 =0.24(9)$.

According to Eq.~\eqref{eq:sigma}, the transition width decreases with acceleration and, at large accelerations, $\sigma$ tends to the constant value $\sigma|_{\alpha \to \infty}\to b_0= 0.80(5)$~fm. This value is of order of the correlation length in gluodynamics, and this might be considered as further confirmation of the smallest transition width.
The fit parameter $b_1$ can be interpreted as the transition width along the temperature axis for very weak acceleration (cf. Eqs.~\eqref{eq:width_rindler},~\eqref{eq:sigma}). In that case, for the considered lattice, $\sigma_T \sim b_1 T_{c0}^2 \simeq 1.6(2)$ MeV. However, this value can not be directly related to the transition width in standard gluodynamics because the effective volumes are different in these cases.
What concerns the susceptibility height \eqref{eq:chi}, it also decreases tending to a constant $d_0=-0.009(24)$ which is zero within the uncertainty. Notice, however, that in the region of strong accelerations our approach might not be applicable and a separate study is required. 

One of the main results of the paper \cite{Chernodub:2024wis}  is the extreme softening of the first order phase transition in standard gluodynamics by acceleration. The authors found that in comparison to the standard gluodynamics, the height $\chi^{max}$ is much smaller and the width $\sigma$ is much larger in the accelerated gluodynamics. In our study, we have observed similar phenomenon in the Rindler spacetime. We believe that it can be explained as follows.
We have already mentioned that the scaling properties of the width and height are $\sigma_T \sim 1/V$ and $\chi^{max}_T \sim V$. In the standard gluodynamics, one substitutes the total volume to this formulas and obtains small width and large height. In the Rindler spacetime, instead of the total volume, one has to use effective volume which is much smaller than the total one. This is because due to TE law only part of the volume has the temperature sufficiently close to the  $T_c$, i.e. in the critical region. 
As a result the width is larger and the height is smaller as compared to standard gluodynamics, leading to the extreme softening. 

\begin{figure*}[t]
    \centering
    \includegraphics[width=0.49\linewidth]{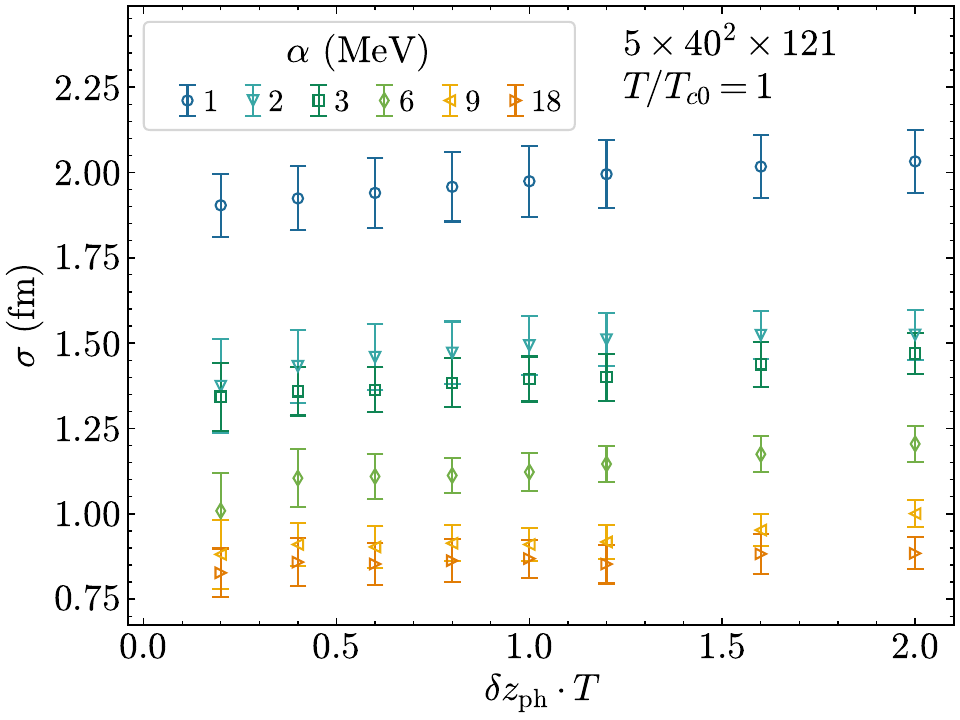}
    \hfill
    \includegraphics[width=0.49\linewidth]{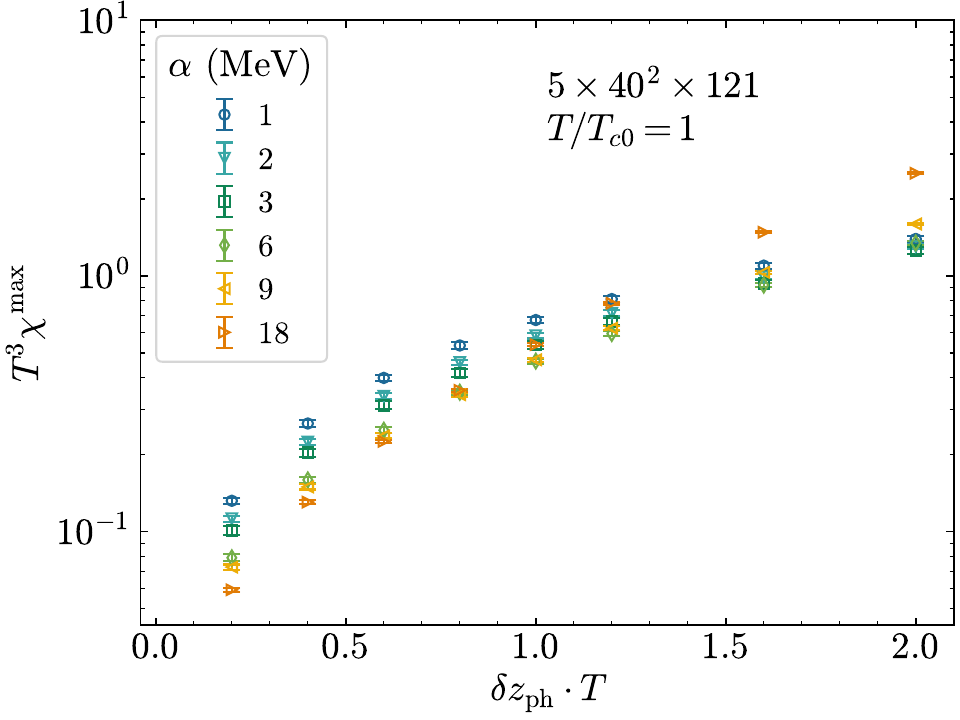}
    \caption{%
    The width $\sigma$ (left) and height $\chi^\text{max}$ (right) of the susceptibility peaks as a function of $\delta z$.
    }
    \label{fig:hw_vs_dz}
\end{figure*}

The results presented in this section were obtained for $\delta z =1$. In addition, we calculated all the observables for various values of $\delta z$, ranging from 1  to $2\times N_t$. Most of our results weakly depend on $\delta z$ in this interval. The only observable that strongly depends on the thickness $\delta z$  is the height of the susceptibility. In Fig.~\ref{fig:hw_vs_dz},  we show the width and the height of the susceptibility as a function of $\delta z$.  
For sufficiently large $\delta z$, the height increases with $\alpha$.
We also find that the mild dependence of $\chi^{\rm max}$ on the volume (shown in Fig.~\ref{fig:hw_vs_volume}, right panel) occurs only for small values of the thickness, as long as it remains much smaller than the transition width.
Starting from a certain value, the height $\chi^{\rm max}$ demonstrates more pronounced behavior with the transverse volume; however, we believe that this is an averaging artifact.

\section{Conclusions}\label{sec:Conclusions}

In this work we investigate the influence of acceleration on the confinement-deconfinement phase transition in gluodynamics. This study is carried out in the comoving reference frame of uniformly accelerated observer which is parameterized by the Rindler coordinates.
In this frame gravitational field appears and the problem is reduced to the study of gluodynamics in external gravitational background. In order to account strongly interacting nature of gluodynamics  we apply the first-principles based, non-perturbative technique of lattice simulations.
In this work we restrict our consideration to the case of small accelerations, i.e. the values much smaller than the inverse correlation length in gluodynamics. Moreover, the volume under study is assumed to be located far from the Rindler horizon. 

We measured the local Polyakov loop, that is the order parameter of the confinement-deconfinement phase transition.
Using Tolman-Ehrenfest law we conducted renormalization of the Polyakov loop in the Rindler spacetime.
Due to the Tolman-Ehrenfest law, the system is effectively cooled down along the direction of acceleration, leading to  coexistence of a confinement phase in the direction of acceleration and a deconfinement phase in the opposite direction. Thus in our study we observe  spatial confinement-deconfinement transition in gluodynamics in the Rindler spacetime. 

Using the inflection point of the local Polyakov loop,  we calculated the position of the boundary between phases~--- the critical distance~--- for various temperatures and accelerations. It was shown that the behavior of the critical distance with temperature and acceleration is in a good agreement with the Tolman-Ehrenfest law, though we observe small deviations. To describe this deviation we included quadratic correction to the Tolman-Ehrenfest law in fitting procedure and determined the parameters of this approximation.

In order to study the properties of the spatial confinement-deconfinement phase transition we measured the susceptibility of the local  Polyakov loop, which reaches the maximum value at the critical distance. Fitting the susceptibility in the vicinity of the critical distance, we determined the width and the height of this observable for various lattice parameters. Our data indicate that in the infinite volume limit the width and the height of the susceptibility remain finite. From this, we drew a conclusion that the spatial confinement-deconfinement transition in the Rindler space is a crossover in agreement with the observation of Ref.~\cite{Chernodub:2024wis}.

Based on our lattice data we concluded that there exists a minimal transition width, which is of order of the correlation length in gluodynamics. In other words the transition from confinement to deconfinement phase takes some spatial distance and it cannot be made smaller than certain value. The hypothesis of the smallest transition width allows one to understand why the spatial phase transition cannot be of the first or second order. Notice that this hypothesis was also confirmed in rotating gluodynamics where it was found that the spatial confinement-deconfinement transition is a crossover~\cite{Braguta:2023iyx, Braguta:2024zpi}. 

In our work we focused on accelerated gluodynamics but we believe that qualitatively our results are valid for QCD. In particular, one can expect that the finite temperature confinement-deconfinement crossover in QCD becomes spatial crossover. In addition, the finite temperature chiral symmetry breaking-restoration crossover of QCD at finite quark masses turns to the spatial crossover. Furthermore, it is reasonable to assume that  at weak accelerations the critical distances of both transitions coincide and are to some accuracy determined by the TE law. However, for sufficiently large accelerations the splitting of the transitions is possible~\cite{Yang:2024tfc}. 

In our work, we mostly focused on the question how acceleration influences the properties of gluodynamics at finite temperature. However, our approach is applicable and the results are valid  in the vicinity of the Schwarzschild black hole horizon.
In particular, our results suggest that at some distance to the horizon spatial confinement-deconfinement phase transition takes place. Consequently, for the radius smaller than the critical one, the vacuum of gluodynamics turns into gluon plasma.\footnote{We believe that similar physical picture can be anticipated in  QCD. One can expect that in QCD at some distances to the horizon the spatial confinement-deconfinement and chiral symmetry breaking-restoration transitions take place and below this distance QCD vacuum becomes quark-gluon plasma. This physical picture is in agreement with S.W.~Hawking's paper \cite{Hawking:1980ng}.}
If one  further approaches to the horizon, at some radius the local acceleration  reaches the values  of order of  the inverse correlation length of gluodynamics or becomes even larger. In this region our results are not valid and an additional investigation of gluodynamics under the influence of these extreme external conditions is necessary. We are going to conduct this investigation in future.  

\begin{acknowledgments}
The authors are grateful to E.T.~Akhmedov, M.N.~Chernodub, O.V.~Teryaev,  E.E.~Kolomeitsev  and V.I.~Zakharov for useful discussions.
This work has been carried out using computing resources of the Federal collective usage center Complex for Simulation and Data Processing for Mega-science Facilities at NRC ``Kurchatov Institute'', http://ckp.nrcki.ru/, and
the heterogeneous computing platform HybriLIT (LIT, JINR)~\cite{hlit:2025}.
This work was supported by the Russian Science Foundation (project no. 23-12-00072-P). 
\end{acknowledgments}

\appendix
\section{Properties of gluon action in the Rindler spacetime} 
\label{app:gluon_action}

The gluon action for the observer located at $z=0$ can be written as  
\begin{equation}
 S_E =  \frac 1 {4 g_{YM}^2} \int d^3  x \int_0^{1/T_0} d \tau  \left ( \frac 1  {1 +{\alpha} z} {({\bs E}^{a} )^2} + (1+{\alpha} z) ({\bs H}^{a} )^2
 \right ) \,. 
 \label{eq:SE_continuum_app}
\end{equation}
As was explained in the Section~\ref{sec:Theory_Rindler}, one can study gluodynamics from the perspective of the other observer located, say, at the $z=z_1$ coordinate.  In this case the Rindler metric~(\ref{eq:metric_z1}) persists  but the acceleration and the temperature should be changed according to the formulas~(\ref{eq:acceleration_z1}) and~(\ref{eq:temperature_z1}).  Similarly to the Rindler metric, one can change the observer in the gluon action. This can be done as follows. First, one passes from the acceleration $\alpha$ in the factor $1+\alpha z$ to the acceleration $\alpha(z_1)$ defined in Eq.~\eqref{eq:acceleration_z1},
\begin{equation}
1+\alpha z = \left ( 1+\alpha z_1 \right ) \left ( 1+ \alpha(z_1) (z-z_1) \right ) \, .
\end{equation}
The constant factor $1+\alpha z_1$ can be absorbed by  new Euclidean time $d \tau_1 = (1+\alpha z_1) d \tau$ in both terms in the gluon action (\ref{eq:SE_continuum_app}) that gives rise to the temperature $T(z_1)$ given by Eq.~(\ref{eq:temperature_z1}). After these manipulations, the gluon action looks as follows
\begin{equation}
 S_E =  \frac 1 {4 g_{YM}^2} \int d^3  x \int_0^{1/T(z_1)} d \tau_1  \left ( \frac 1 {(1+\alpha z_1)^2} \frac 1  {1 +\alpha(z_1)(z-z_1) } {({\bs E}^{a} )^2} + \left (  1 +\alpha(z_1)(z-z_1)  \right ) ({\bs H}^{a} )^2 \right ) \,.
\end{equation}
One can get rid of the redundant factor $1/(1+\alpha z_1)^2$ in last expression, if in the chromoelectic field $\bs E = \partial_4 \bs A -  \bs \nabla A_4 + [A_4, \bs A]$ one passes to new variables $ \partial_4  \to (1+\alpha z_1) \partial_4$ and $A_4 \to (1+\alpha z_1) A_4$. 
As a result the following expression can be obtained,
\begin{equation}
S_E = \frac 1 {4 g_{YM}^2} \int d^3 x \int_0^{1/T(z_1)} d \tau_1  \left ( \frac 1  {1 +{\alpha(z_1)} (z-z_1)}  {\left({\bs E}^{a} \right)^2} + \left (1+{\alpha(z_1)} (z-z_1) \right ) \left({\bs H}^{a} \right)^2
\right ) \,,
\label{eq:SE_continuum_z1_app}
\end{equation}
which is exactly Eq.~\eqref{eq:SE_continuum_z1}.
Thus one sees that the gluon action (\ref{eq:SE_continuum_app}) has the same form for an arbitrary observer if the acceleration and the temperature is taken at the corresponding position (compare Eqs.~\eqref{eq:SE_continuum_app} and~\eqref{eq:SE_continuum_z1_app}). 
For some theories the action in the Rindler metric may be simplified even further.
For example, let us consider the action of the scalar field $\varphi$ in the Rindler space 
\begin{equation}
 S_E =  \frac 1 {2} \int d^3  x \int_0^{1/T_0} d \tau  \left ( \frac 1  {1 +{\alpha} z} (\partial_4 \varphi)^2 + (1+{\alpha} z) \left ( 
 (\bs \nabla \varphi)^2 + m^2 \varphi^2
 \right ) \right ) \,,
 \label{eq:scalar_rindler_app}
\end{equation}
and apply the trick used above. In this case absorbing the position dependent factor into the Euclidean time, $d\tau \to (1+\alpha z) d\tau$, and changing variable in the action, $\partial_4 \to (1+ \alpha z) \partial_4$, we get
\begin{equation}
 S_E =  \frac 1 {2} \int d^3  x \int_0^{1/T(z)} d\tau  \left (  (\partial_4 \varphi)^2 +   
 (\bs \nabla \varphi)^2 + m^2 \varphi^2
 \right ) \,. 
 \label{eq:scalar_rindler_Tz_app}
\end{equation}
The last expression shows that, after changing variables, the action of the scalar field in the Rindler space exactly coincides with the action in flat space. So we conclude that the influence of the acceleration to scalar field is {\it exactly} reduced to the variation of the temperature according to the formula  $T(z)=T_0/(1+\alpha z)$ which is Tolman-Ehrenfest law~\cite{Tolman:1930ona}. 
Notice also that the last expression Eq.~\eqref{eq:scalar_rindler_Tz_app}, is in agreement with Luttinger relations, which connects acceleration to the gradient of temperature~\cite{Luttinger:1964zz}.

Similarly to the scalar field, one can try to eliminate the coordinate dependent factor $1+ \alpha z$ in the gluon action. However, this cannot be done in this case since changing variables $A_4 \to (1+\alpha z) A_4$ does not commute with the derivative $\partial_z A_4$ in the $z$-component of the chromoelectric field operator, $E_z = \partial_4 A_z -  \partial_z A_4 + [A_4, A_z]$, and violate gauge invariance. We conclude that gluodynamics in the Rindler space cannot be {\it exactly} reduced to the gluodynamics with varying temperature in the flat space. Nevertheless, one can apply the transformation $A_4 \to (1+\alpha z) A_4$, $d\tau \to (1+\alpha z) d\tau$  and for weak acceleration get the following form of the action 
\begin{equation}
S_E=S_0 + \mathcal{O} (\alpha)\,,
\end{equation} 
where 
\begin{align} 
S_0&= \frac 1 {4 g_{YM}^2} \int d^3  x \int_0^{1/T(z)} d \tau  \left ( ({\bs H}^{a} )^2 + {({\bs E}^{a} )^2} \right )\,. 
\end{align}
The leading term $S_0$ is the gluon action in the flat space with spatial dependence of the temperature according to the TE law.
The corrections to the leading order approximation are determined by higher order terms which are designated as $\mathcal{O}(\alpha)$. In particular, the first order correction to $S_0$ is given by the operator $\sim \alpha E_z^a A_4^a$, whereas the second order correction is determined by the operator $\sim \alpha^2 (A_4^a)^2$.
It is interesting to note that the authors of paper~\cite{Chernodub:2024wis} studied accelerated gluodynamics using the leading term $S_0$. From our consideration it is clear that the study of accelerated gluodynamics in the Rindler spacetime differs from that in paper~\cite{Chernodub:2024wis} by higher order corrections which might become noticeable for sufficiently large acceleration.

\section{Renormalization of the Polyakov loop in the Rindler spacetime}\label{app:Renormalization}

To renormalize the Polyakov loop, Eq.~\eqref{eq:L_z_ren}, we adapt the multiplicative renormalization scheme from the standard gluodynamics~\cite{Kaczmarek:2002mc, Gupta:2007ax} to the Rindler spacetime.
In the Euclidean space the Polyakov loop is renormalized as follows,
\begin{equation}
\label{eq:renormalization}
    L = \exp\left( \log Z(g^2)\cdot \frac{L_t}{a(g^2)}\right) L^b = \left(Z(g^2)\right)^{N_t} L^b\,,
\end{equation}
where $a$ is the lattice spacing and $L_t = N_t a = 1/T$ is the size of the lattice in the compactified Euclidean time direction.
In Appendix~\ref{app:gluon_action} it was shown  that for weak accelerations gluodynamics in the Rindler spacetime is reduced to the standard gluodynamics with the local temperature determined by the TE law $T(z) = T/(1 + \alpha z)$.
So, in the Rindler spacetime it is reasonable to apply the renormalization prescription~(\ref{eq:renormalization}) with  $L_t$ replaced by  $L_t(z) = 1/T(z)$, which takes the following form:
\begin{equation}
\label{eq:renormalization1}
    L(z) = \exp\left( \log Z(g^2)\cdot N_t (1 + \alpha z) \right) L^b = \big(Z(g^2) \big)^{N_t (1+\alpha z)} L^b(z)\,.
\end{equation}

\begin{figure*}[t]
    \centering
    \includegraphics[width=0.49\linewidth]{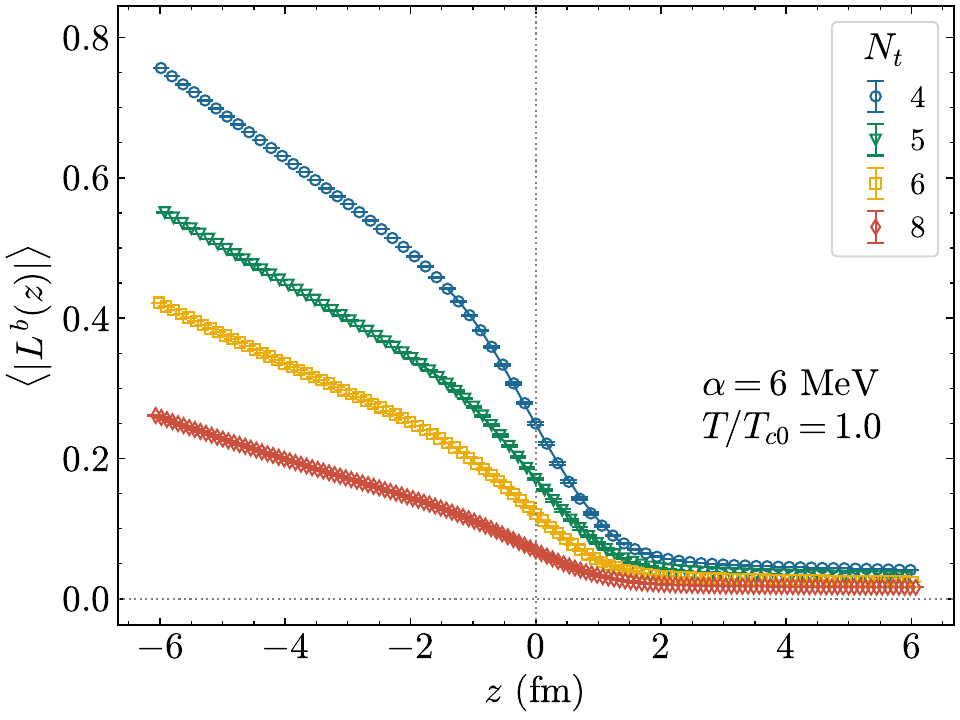}
    \hfill
    \includegraphics[width=0.49\linewidth]{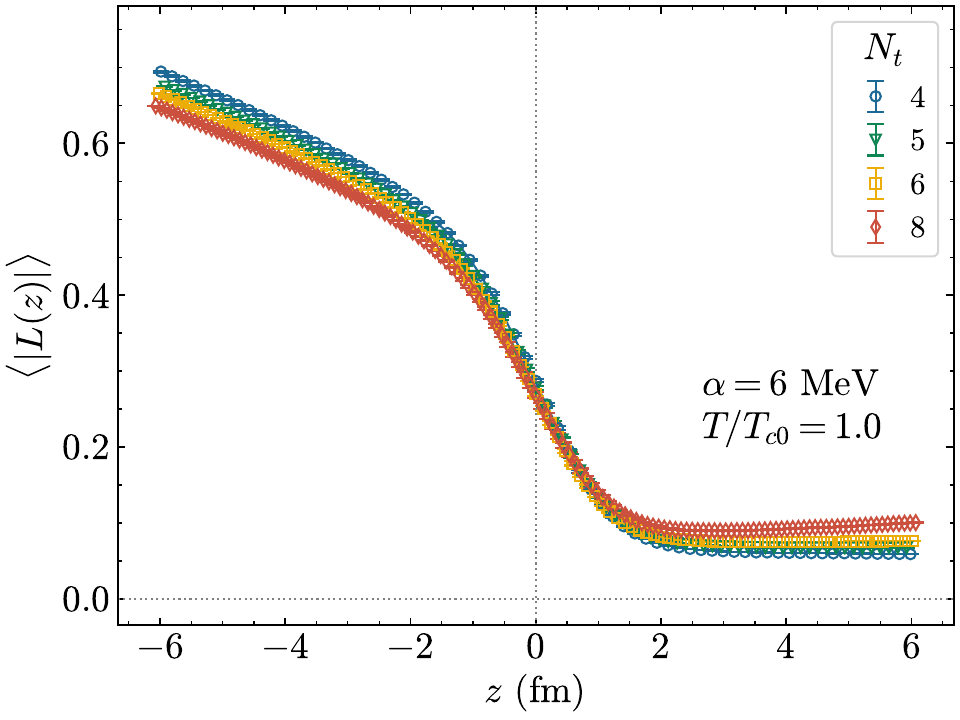}
    \caption{%
    The bare (left panel) and renormalized (right panel) local Polyakov loop $L(z)$ as a function of $z$. The results were obtained on the lattices with different $N_t = 4,\, 5,\, 6,\, 8$ and fixed ratios $N_s/N_t = 8$, $(N_z-1)/N_t = 24$ at $\alpha = 6$~MeV.
    }
    \label{fig:Lz_compareNt_ren}
\end{figure*}

Figure~\ref{fig:Lz_compareNt_ren} demonstrates validity of the renormalization scheme at $T=T_{c0}$ and $\alpha=6$~MeV for all studied lattice sizes of $N_t=4,\, 5,\, 6,\, 8$ with common aspect ratio $N_s/N_t = 8$ and $(N_z -1)/N_t = 24$. The left and right panels depict the bare and renormalized (modulus of the) Polyakov loop as a function of $z$ coordinate, respectively. The $N_t$ dependence is significantly reduced in the renormalized Polyakov loop.
We believe that the minor cut-off dependence at low $z$ may be explained by proximity to the critical temperature.
In standard homogeneous gluodynamics, the renormalized Polyakov loop is also subjected by more noticeable cut-off effects in the vicinity of critical temperature (see Ref.~\cite{Lo:2013hla}). 
In addition, the renormalization factor $Z(\beta)$ is well computed in the deconfinement phase and has large systematic uncertainties near the transition point (cf. results for renormalization factor from lattices with $N_t =4$ and $N_t = 8$ around $\beta_c$ in Ref.~\cite{Gupta:2007ax}). At large $z$, system is in the confinement phase, and small non-zero values of the Polyakov loop are attributed to the finite volume effects (see also Fig.~\ref{fig:Lz_vs_volume}~(left)). In the infinite volume limit the modulus of the Polyakov loop is zero in confinement phase for all $N_t$.

\begin{figure*}[t]
    \centering
    \includegraphics[width=0.49\linewidth]{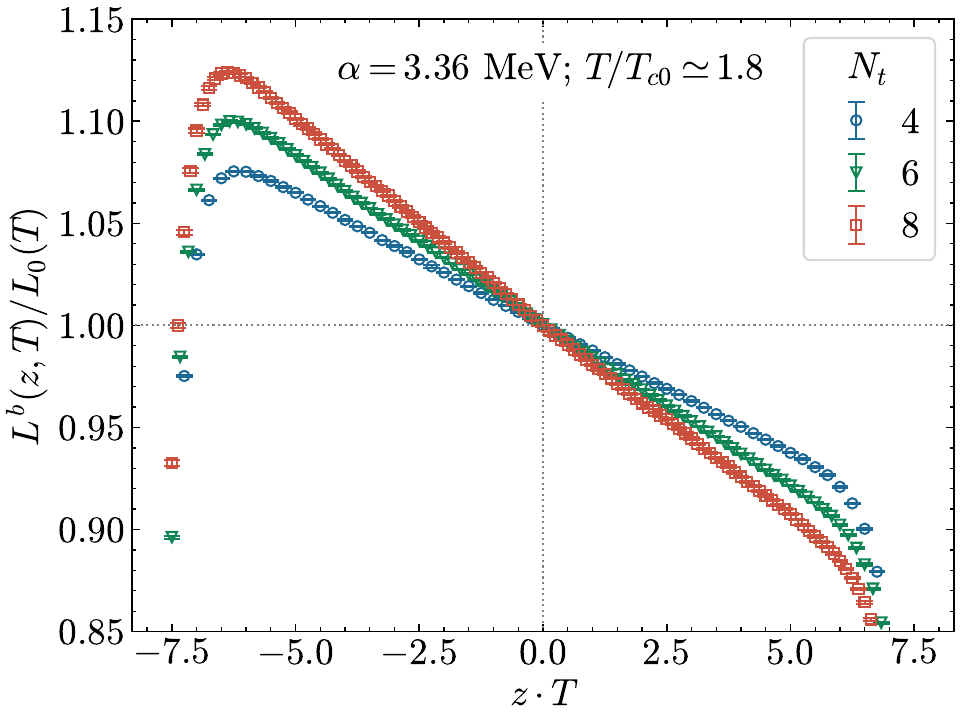}
    \hfill
    \includegraphics[width=0.49\linewidth]{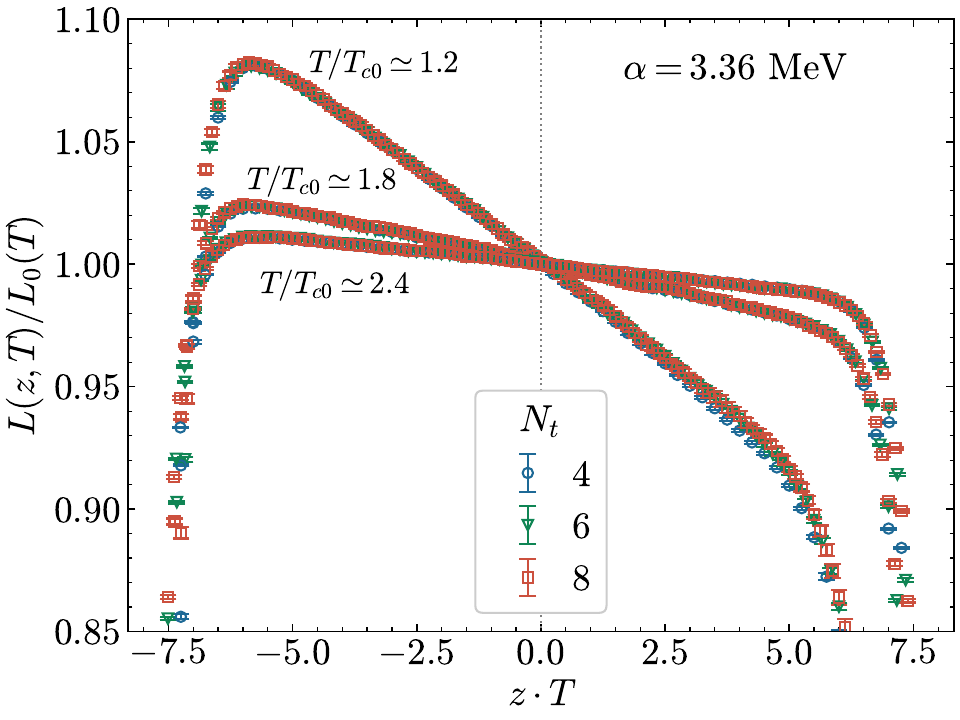}
    \caption{%
    (left) The ratio of bare Polyakov loop of accelerated system to that of the standard homogeneous system for $N_t=4, 6, 8$ at $\alpha = 3.36$~MeV and $T/T_{c0}=1.8$.
    (right) The ratio of renormalized (modulus of) Polyakov loop of accelerated system with that of the standard homogeneous systems for $N_t=4, 6, 8$ and $N_s/N_t = 8$,~$(N_z -1)/N_t = 24$ at $T/T_{c0} = 1.2, 1.8, 2.4$ and $\alpha=3.36$~MeV.
    }
    \label{fig:ren_polyakov}
\end{figure*}

In Fig.~\ref{fig:ren_polyakov}, we compare renormalization in the Rindler spacetime with that in the standard gluodynamics. In contrast to Fig.~\ref{fig:Lz_compareNt_ren} here we also show the points affected by open boundary condition (OBC).
The left panel of Fig.~\ref{fig:ren_polyakov} shows the ratio of bare (modulus of) Polyakov loop in the Rindler spacetime to that in the standard gluodynamics as a function of $z$. If one conducts the renormalization of the both Polyakov loops as this is done in the standard gluodynamics~\eqref{eq:renormalization}, the renormalization factor  cancels and one gets the ratio depicted on the left panel of Fig.~\ref{fig:ren_polyakov}. However, one sees a clear discrepancy between the Polyakov loops for different $N_t$, i.e. additional $z$-dependent renormalization is necessary. 
The renormalization~\eqref{eq:renormalization1} in the Rindler spacetime is shown on the right panel of Fig.~\ref{fig:ren_polyakov}, where we plot the ratio of the renormalized Polyakov loops in both theories. In addition to the temperature $T/T_{c0}=1.8$ shown on the left panel, we present these ratios at the temperatures $T/T_{c0}=1.2$ and $T/T_{c0}=2.4$.  An excellent agreement between the results obtained at different $N_t$ is seen.   

\section{Finite volume effects on the critical distance}\label{app:finite_volume}

\begin{figure*}[t]
    \centering
    \includegraphics[width=0.49\linewidth]{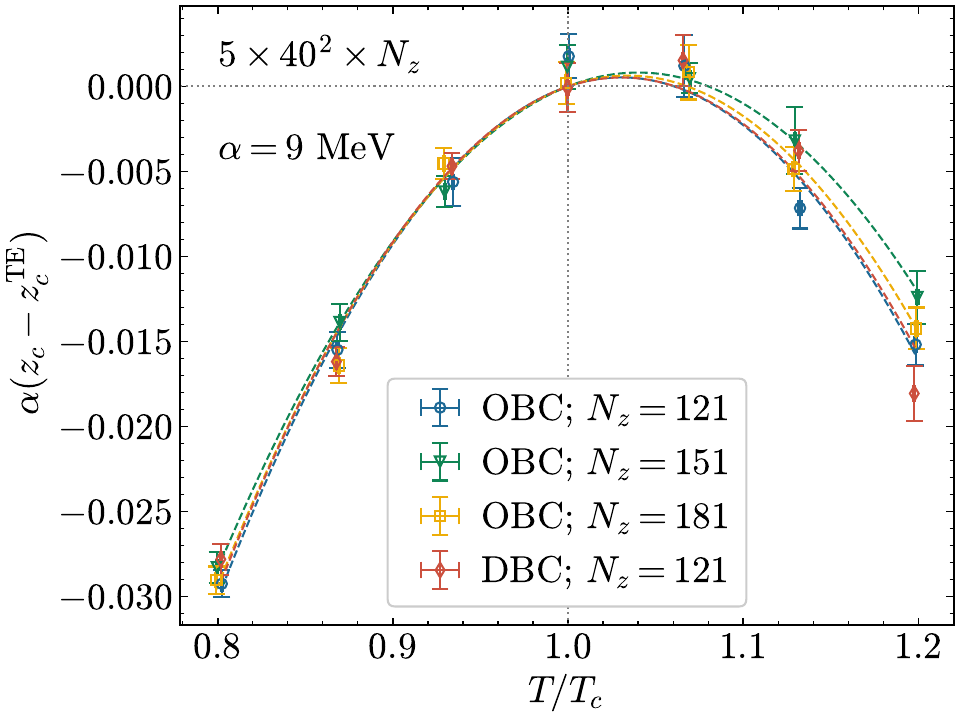}
    \hfill
    \includegraphics[width=0.49\linewidth]{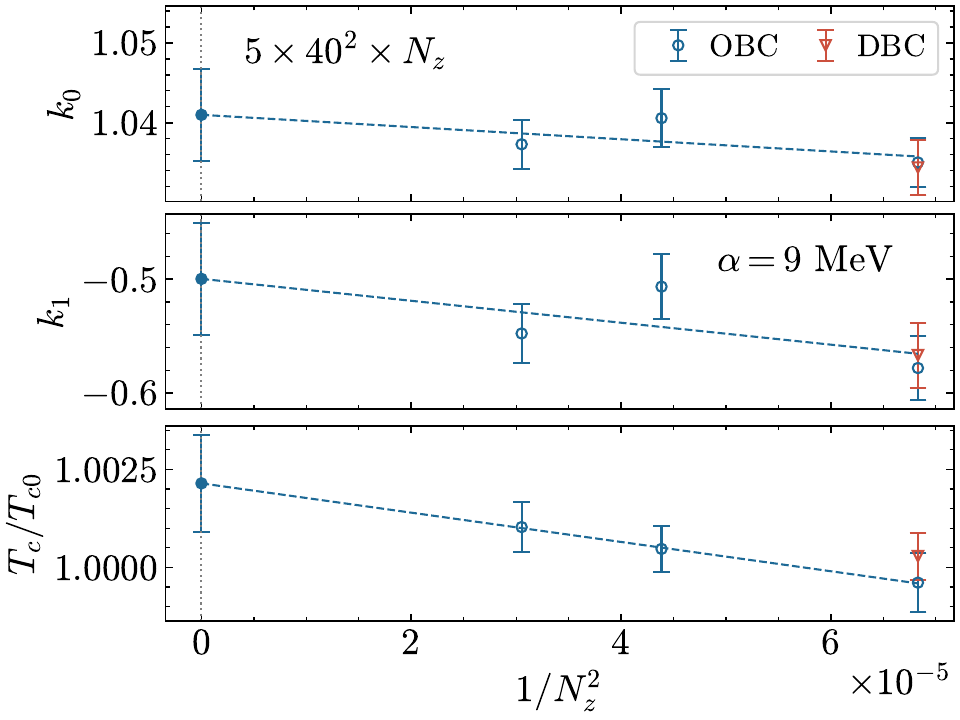}
    \caption{%
    (left) The difference between $z_c$ and $z_c^{\text{TE}}$ as a function of temperature for the lattices of size $5 \times 40^2 \times N_z$ with different $N_z = 121, 151, 181$ for open and Dirichlet boundary conditions at acceleration $\alpha = 9$~MeV.
    (right) The best fit parameters $k_0$, $k_1$ and $T_c$ as a function of $N_z$ for the  same lattices at acceleration $\alpha = 9$~MeV. 
    }
    \label{fig:zc_vs_T_testNz}
\end{figure*}

In Section~\ref{sec:Results_TE_checking}, we presented the results for the lattices with fixed aspect ratios, $N_s/N_t = 8$ and $(N_z-1)/N_t = 24$. 
To estimate the influence of finite longitudinal lattice size on our results, we perform the calculations on the lattices $5\times 40^2\times N_z$ with $N_z = 121, 151, 181$ at acceleration $\alpha = 9$~MeV. 
In addition, to study the effect of boundary conditions we conduct  lattice simulation with Dirichlet boundary condition (DBC) in $z$-direction.
For this lattice set up we calculate the local Polyakov loop and fit it by the function~\eqref{eq:fit_tanh} to find the critical distance, $z_c$.
The critical distance as a function of temperature is presented in Fig.~\ref{fig:zc_vs_T_testNz}.
One can see that the results for different $N_z$ are in a good agreement with each other and that obtained for DBC, which suggests that the effects of open boundary conditions implemented in the longitudinal $z$-direction are under control in our study. Furthermore,  the quadratic deviation from the predictions of the TE law is not a consequence of the specific choice of boundary conditions.

We fit the critical distance by Eq.~\eqref{eq:zc_fit}, in the same way as we did in Section~\ref{sec:Results_TE_checking}.
The fit parameters $k_0$, $k_1$, $T_c$ as a function of $N_z$ are shown in Fig.~\ref{fig:zc_vs_T_testNz} (right).  It is seen that they exhibit a very mild dependency on $N_z$.
Note that one cannot extend the system to infinity in the $z<0$ direction, as the horizon, $z_h = -1/\alpha$ should lie outside the studied volume.

\begin{figure*}[t]
    \centering
    \includegraphics[width=0.49\linewidth]{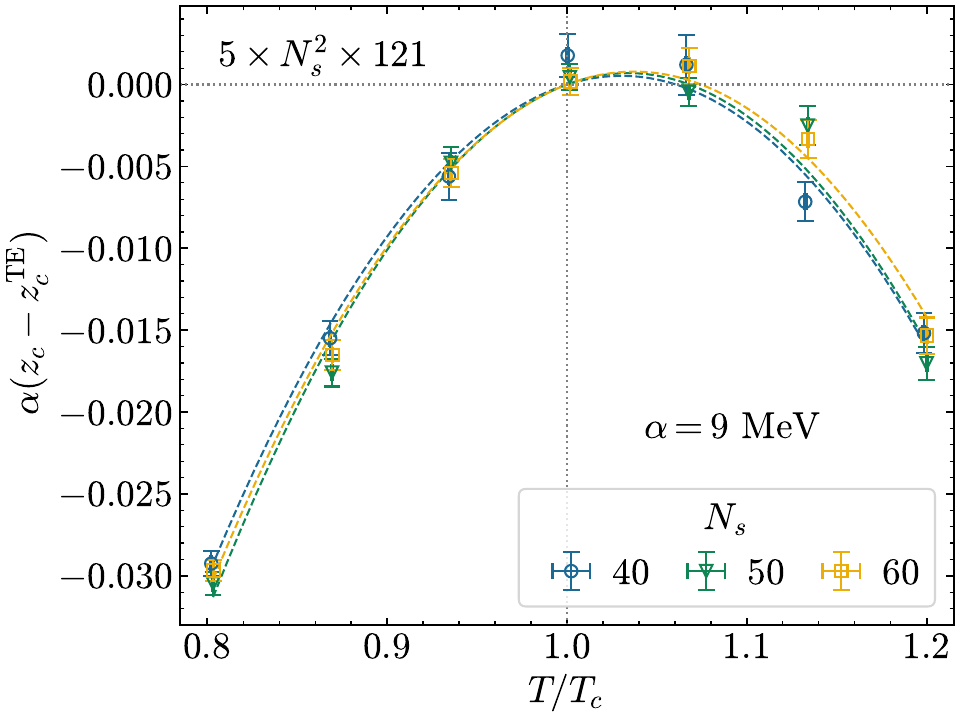}
    \hfill
    \includegraphics[width=0.49\linewidth]{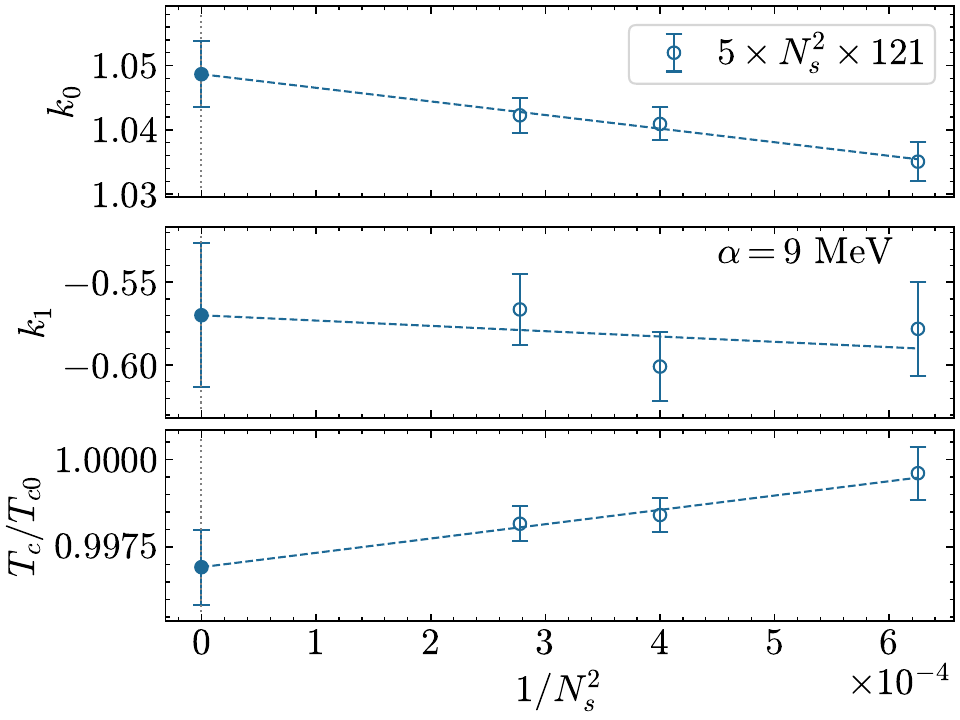}
    \caption{%
    (left) The difference between $z_c$ and $z_c^{\text{TE}}$ as a function of temperature for the lattices of size $5 \times N_s^2 \times 121$ with different $N_s = 40, 50, 60$.
    (right) The best fit parameters $k_0$, $k_1$ and $T_c$ as a function of $N_z$ for the same lattices at acceleration $\alpha = 9$~MeV. 
    }
    \label{fig:zc_vs_T_testNs}
\end{figure*}

On the contrary, the infinite volume limit may be taken in the transversal directions. As already discussed in Section~\ref{sec:Results_Crossover}, the position of the crossover at $T=T_{c0}$ shifts slightly, when the transversal system size, $N_s$, is increased.
To analyze how the finite volume affects the fit coefficients and deviation from the TE law, we calculate the critical distances for lattices $5\times N_s^2\times 121$ with $N_s = 40, 50, 60$ at several temperatures (see Fig.~\ref{fig:zc_vs_T_testNs} (left)) and fit these data by the function~\eqref{eq:zc_fit}.

One can see that the critical temperature $T_c$ drops with the transversal size only by $\sim 0.3\%$, and this trend is opposite to the one from Fig.~\ref{fig:zc_vs_T_testNz}. This minor shift does not affect the coefficients, $k_0$ and $k_1$, significantly.
Figure~\ref{fig:zc_vs_T_testNs} (right) shows the values of the fit coefficients and the infinite volume extrapolated results.
From these results we conclude that the finite volume effects in the coefficient $k_0$ do not exceed $\sim1\%$, and quadratic deviation from the predictions of TE law persists even in the infinite volume limit, $N_s \to \infty$.

\section{The critical distance extracted from different observables}\label{app:chi_vs_L_vs_bare}

\begin{figure*}[t]
    \centering
    \includegraphics[width=0.49\linewidth]{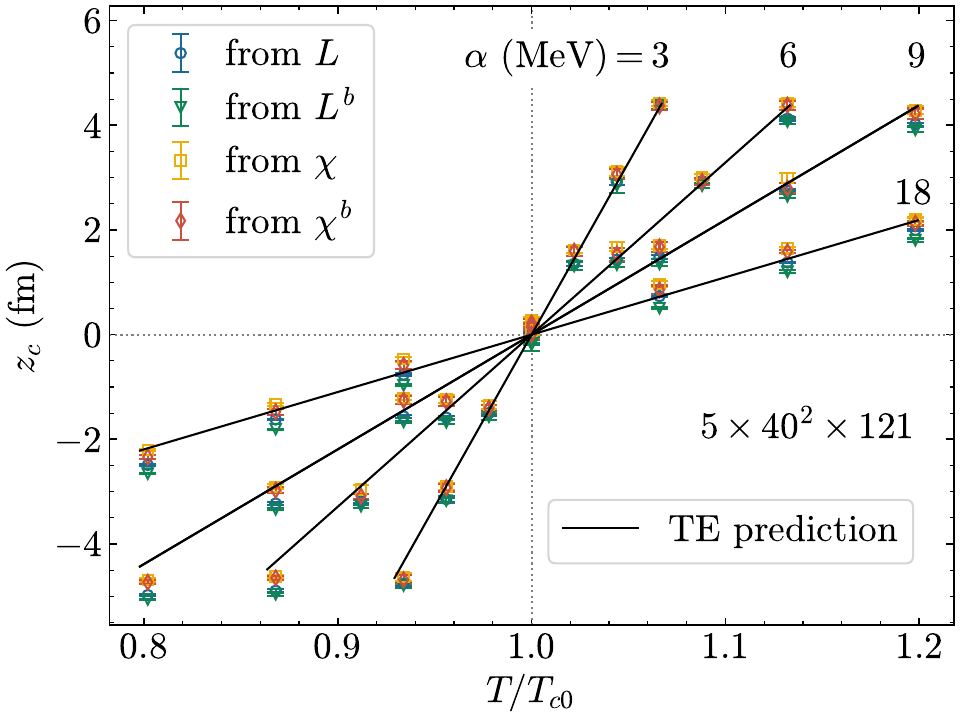} \hfill
    \includegraphics[width=0.49\linewidth]{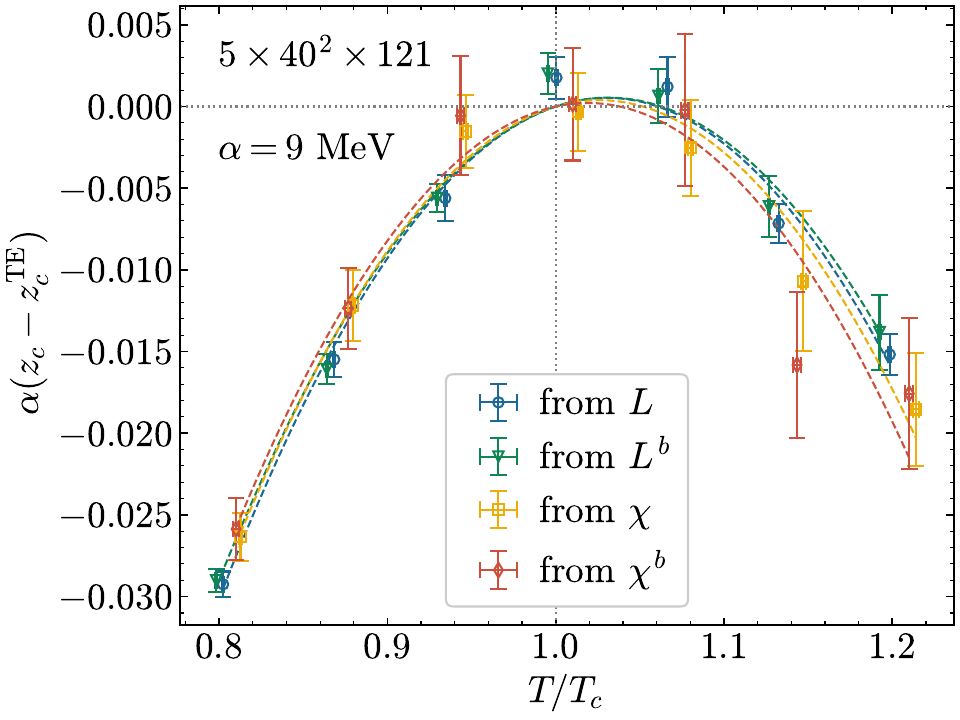}
    \caption{%
    (left) The critical distances $z_c$ as a functions of temperature.
    (right) The difference between $z_c$ and $z_c^{\text{TE}}$ as a function of temperature.
    The data are obtained  on the lattice $5 \times 40^2 \times 121$ at $\alpha=9$ MeV from the following observables: bare~($L^b$) and renormalized~($L$) Polyakov loop, bare~($\chi^b$) and renormalized~($\chi$) Polyakov loop susceptibility.
    }
    \label{fig:zc_definitions_plA_chiA}
\end{figure*}

As discussed in Section~\ref{sec:Results_Crossover}, the spatial transition in accelerated gluodynamics exhibits crossover behavior. This circumstance implies that various observables can be used to determine the critical distance, and, in general, different definitions of $z_c$ do not necessarily coincide.
Using the methods introduced in Section~\ref{sec:Observables}, we calculated the critical distance from the inflection point of the local Polyakov loop, as well as from the position of susceptibility peak, to compare the results.
Figure~\ref{fig:zc_definitions_plA_chiA} (left) represents the critical distances, determined from different observables on the lattice $5\times 40^2\times 121$, as a function of temperature at several accelerations.
Here we used both renormalized and bare operators.
The figure shows that different definitions yield slightly shifted results.
For most lattice parameters, the critical distance from the susceptibility is larger than that from Polyakov loop by about~$\sim 0.1-0.4$~fm, i.e. it is shifted in the direction of acceleration.
Nevertheless, the results have the same trend with temperature.

We fit all sets of data by the function~\eqref{eq:zc_fit} and find the best fit values of the parameters $k_0$, $k_1$ and $T_c$.
As an example, the rescaled difference, $\alpha (z_c - z_c^{\rm TE})$, is shown in Fig.~\ref{fig:zc_definitions_plA_chiA} (right) as a function of $T/T_{c}$ at acceleration $\alpha = 9$~MeV.
One can see that the data are in an excellent agreement in these coordinates.
In this figure, a minor shift of the data points along the temperature axis arises from the best fit values of $T_c$ used for normalization.
At this acceleration, we obtain $T_c^{({\rm from}\,L)}/T_{c0} = 0.9996(8)$, $T_c^{({\rm from}\,\chi)}/T_{c0} = 0.9869(14)$ from the renormalized operators, and 
$T_c^{({\rm from}\,L^b)}/T_{c0} = 1.0047(6)$, $T_c^{({\rm from}\,\chi^b)}/T_{c0} = 0.9900(20)$ from the bare operators.
A similar  consistency in the rescaled deviation from the TE law was found for all lattice parameters used in this study. 
Reasonable agreement between the results for bare and renormalized data is observed due to a smooth behavior of the renormalization function~\eqref{eq:L_z_ren}, but it may worsen at large accelerations.

We also checked how the results depend on the thickness $\delta z$ of the averaged volume. 
We calculated continuum limit extrapolated results for the fit parameters with two definitions of critical distance and for several thicknesses $\delta z = 1,\, N_t,\, 2 N_t$. 
We found that the coefficients $k_0$ and $k_1$ are consistent within uncertainties for all values of $\delta z$, and the shift in $T_c$, discussed above is diminished at large $\delta z$.

\bibliographystyle{apsrev4-2}
\bibliography{gravity}

\end{document}